 \documentclass[modern]{aastex631}

\usepackage{rotating}
\usepackage{pgffor}  
\usepackage{mwe}
\usepackage{graphics}
\usepackage{pgffor}

\shorttitle{Low mass members in Rho Oph}
\shortauthors{Sharma et al.}
\graphicspath{{./}{figures/}}
\usepackage{hyperref}

\begin{document}
\author[0000-0003-1634-3158]{Tanvi Sharma}
\affiliation{Institute of Astronomy, National Central University, 300 Zhongda Road, Zhongli 32001 Taoyuan, Taiwan}

\author[0000-0003-0262-272X]{Wen Ping Chen}
\affiliation{Institute of Astronomy, National Central University, 300 Zhongda Road, Zhongli 32001 Taoyuan, Taiwan}
\affiliation{Department of Physics, National Central University, 300 Zhongda Road, Zhongli 32001 Taoyuan, Taiwan}

\author[0000-0003-4614-7035]{Beth Biller}
\affiliation{Institute for Astronomy, Royal Observatory, University of Edinburgh, Blackford Hill, Edinburgh EH93HJ, UK}

\author[0000-0003-0475-9375]{Lo\"ic Albert}
\affiliation{Institut Trottier de recherche sur les exoplan\`etes, Universit\'e de Montr\'eal, Montr\'eal, QC, H3C 3J7, Canada}
\affiliation{D\'epartement de physique, Universit\'e de Montr\'eal, Montr\'eal, QC, H3C 3J7, Canada}

\author[0000-0002-2234-4678]{Belinda Damian}
\affiliation{SUPA, School of Physics \& Astronomy, University of St Andrews, North Haugh, St Andrews, KY16 9SS, UK}
\affiliation{Department of Physics and Electronics, CHRIST (Deemed to be University), Hosur Road, Bengaluru 560029, India}

\author{Jessy Jose}
\affiliation{Department of Physics, Indian Institute of Science Education and Research Tirupati, Yerpedu, Tirupati - 517619, Andhra Pradesh, India}

\author[0000-0003-1618-2921]{Bhavana Lalchand}
\affiliation{Institute of Astronomy, National Central University, 300 Zhongda Road, Zhongli 32001 Taoyuan, Taiwan}

\author[0000-0003-2232-7664]{Michael C. Liu}
\affiliation{Institute for Astronomy, University of Hawai'i, 2680 Woodlawn Drive, Honolulu HI 96822}

\author{Yumiko Oasa}
\affiliation{Faculty of Education / Graduate School of Science and Engineering, Saitama University
255, Shimo-Okubo, Sakura, Saitama 338–8570, Japan}

\title{A Novel Survey for Young Substellar Objects with the W-band Filter. VII. Water-Bearing Objects in the Core of the Rho Ophiuchi Cloud Complex
      }

\begin{abstract}

We present a study of very low-mass stars and brown dwarfs in the rich star-forming core of the Rho Ophiuchi cloud complex. The selection of the sample relies on detecting the inherent water absorption characteristic in young substellar objects. Of the 22 water-bearing candidates selected, 15 have a spectral type of M6 or later.  Brown dwarf candidates too faint for membership determination by Gaia have their proper motions derived by deep-infrared images spanning six years. Astrometric analysis confirms 21/22 sources as members, one identified as a contaminant.  Infrared colors and the spectral energy distribution of each water-bearing candidate are used to diagnose the mass, age, and possible existence of circumstellar dust. 15 sources exhibit evidence of disks in their spectral energy distributions, as late as in M8-type objects. Spectroscopy for bright candidates has confirmed one as an M8 member and verified two sources (with disks) exhibiting signatures of magnetospheric accretion.

\end{abstract}

\section{Introduction}\label{sec:Introduction}

Brown dwarfs (BDs) are often referred to as ``failed stars'' because they do not reach core temperatures high enough to sustain hydrogen fusion like stars.  In terms of formation and evolution, BDs bridge the mass gap between stars and planets, i.e., between $\la0.07~M_\sun$ and $\la0.013~M_\sun$ corresponding to the hydrogen and deuterium burning limit, respectively \citep{1963ApJ...137.1121K, 1997ApJ...491..856B}. Even though there are numerous observations, there is still a lack of clear understanding regarding their formation mechanism, though various scenarios have been proposed.  For example, BDs could form via turbulent fragmentation of clouds \citep{2002ApJ...576..870P,2008MNRAS.389.1556B, 2012MNRAS.419.3115B}; i.e., they could form like stars.  Or, they could form like planets via gravitational instability \citep{2006A&A...458..817W} within young stellar disks.  Alternatively, \citet{2001AJ....122..432R, 2005MNRAS.356.1201B} suggest an ejection scenario in which a protostellar core is expelled from a stellar cluster, thereby ceasing accretion to become a star.    

So far the most widely acknowledged mechanism is that BDs form like low-mass stars, supported by observational evidence of disks and accretion \citep[e.g.,][]{2003AJ....126.1515J, 2004A&A...424..603N,  2008ApJ...672L..49S}.  However, studies by  \citet{2015A&A...579A..66M}, \citet{2017A&A...600A..20A}, and \citet{2023AJ....166..262B} suggest more active accretion in younger substellar objects, whereas no such age effect is observed in the stellar sample.  If so, there must be additional factors affecting substellar formation.  It is thus desirable to study BDs versus low-mass stars to diagnose their possible differences in the earliest evolution.

Young BDs (i.e., with late M to early L spectral types) exhibit prominent absorption due to water molecules \citep{2020PASP..132j4401A}. For even cooler objects (i.e., late L to T types), methane begins to dominate \citep{2015ApJ...811L..16C}.  Spectroscopic identification of these intrinsically faint objects is feasible only if a reliable list of substellar targets is available.  Broadband photometric observations to recognize cool objects often lead to a large fraction of contaminants such as reddened background stars or active galaxies.  An effective way is to detect molecule-bearing objects by on-off imaging with filters centered on water or methane bands.  This work is a part of one such program, called the ``W Band'' survey using near-infrared photometry sensitive to the water absorption near 1.45~$\micron$ as a detection marker to identify young BDs in nearby star-forming regions. The technique and design of the custom-made water filter with implementation and test results are described in \citet{2020PASP..132j4401A}. Subsequent applications have been reported for Serpens South and Serpens Core  \citep{2020ApJ...892..122J, 2021MNRAS.505.4215D, 2023MNRAS.520.3383D}, for Perseus \citep{2022AJ....164..125L}, and for $\sigma$ Orionis \citep{2023ApJ...951..139D}.  Here we present the identification and characterization of BD candidates in Rho Ophiuchi.

The Rho Ophiuchi cloud complex is a nearby star-forming region with filamentary structures \citep{1989ApJ...338..902L, 2020A&A...638A..74L} including L\,1688, with the sub-cloud L\,1688A being the most active in star formation \citep{2008hsf2.book..351W}. The youth \citep[0.3~Myr]{1995ApJ...450..233G, 1999ApJ...525..440L} and proximity \citep[$\sim137.3$~pc]{2017ApJ...834..141O, 2020A&A...633A..51Z} makes it an optimal laboratory to investigate the evolutionary stages of stellar and substellar objects in their infancy. The complex has been the focus of ample studies of young stars, for the substellar population, from discovery \citep{  1999ApJ...525..440L, 2011AJ....142..140E,2010A&A...515A..75A, 2017A&A...597A..90D,2020AJ....159..282E}, to presence of disks around BDs \citep{2016A&A...593A.111T, 2019MNRAS.482..698C}, to observations of accretion and outflows \citep{2006A&A...452..245N, 2022CPhy...32...95N}.  

This work focuses on the main core of the complex, i.e., L\,1688A, using the W-filter to unveil embedded BD members possibly not reported before, for which spectroscopy has been taken for some bright candidates to confirm their BD nature or to detect accretion signature. We also analyzed the spectral energy distribution (SED) or infrared colors of each candidate to probe possible existence of circumstellar dust.  Moreover, proper motions of candidates too faint for Gaia have been derived with deep infrared images to constrain the membership.


\section{Data Analysis and Results} \label{sec:Data}

\begin{figure}[htb!]
    \includegraphics[width=\columnwidth]{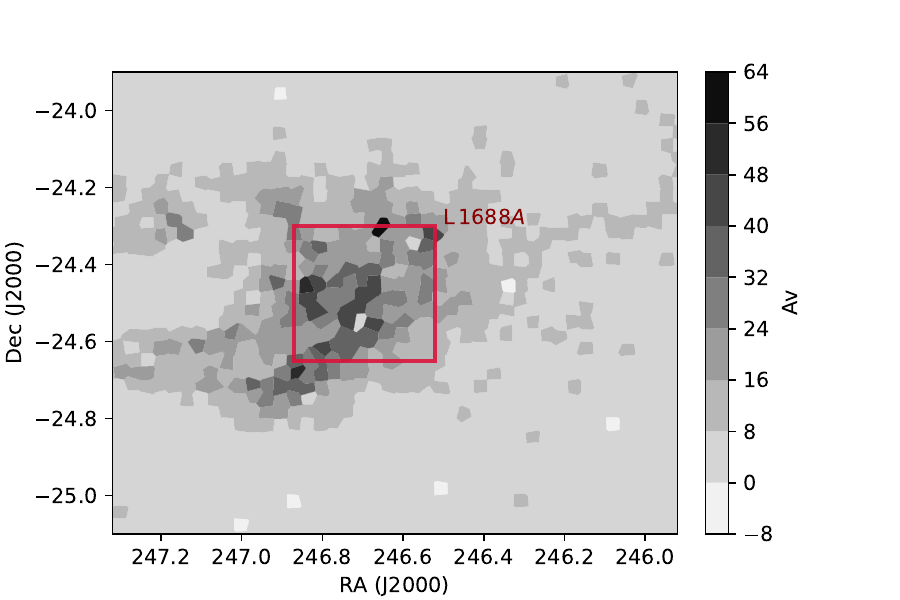} %
    \caption{The spatial coverage of our study of L\,1688A, with the underlying extinction map being computed using 2MASS catalog \citep{2003yCat.2246....0C} adopting the NICE technique from \citet{1994ApJ...429..694L}.  The color bar shows the extinction range.  The red square encloses the area of the W~band observations covering a field of $20\arcmin\times 20\arcmin$ corresponding  to a linear scale of 0.8 pc.   
    }%
    \label{fov_data}
\end{figure}

Figure~\ref{fov_data} illustrates the extent of our studied region.  Sec.~\ref{sec:w_band data} describes the W-band images used to identify BD candidates.  Sec.~\ref{sec:astrometric data} then reports how we used archival images to derive the proper motion of faint candidates in reference to the Gaia DR3 measurements of bright sources.  The SED and color analyses with multiwavelength photometric data are presented in Sec.~\ref{sec:multi_wav data}.  Sec.~\ref{sec:spectroscopy obs} describes confirmation infrared spectroscopic observations. 

\subsection{W-Band Imaging and Selection of Candidates}\label{sec:w_band data}

We used imaging observations from WIRCam implementing the custom-made water band filter (i.e., W-band) mounted on the CFHT 3.6~m telescope. WIRCam covered a $\sim 20\arcmin \times 20\arcmin$ field of view with a pixel scale of $0\farcs3$ \citep{2004SPIE.5492..978P}. The observations were carried out on 2016 March 18 and July 20, with each epoch having 4 images per filter in $J$ (1.25~microns), $H$ (1.65~microns), and $W$ (1.45~microns) bands, constituting a total integration time of 80~s in the $J$ and $H$ bands, and 520~s in the $W$ band. The images were processed by the I'iwi pipeline. The photometry in $J$ and $H$ bands was calibrated to 2MASS. The W-band calibration was conducted according to the procedure described in \citet{2020PASP..132j4401A}.  The WIRCAM images contained 313 sources.  

To diagnose the possible water absorption, \citet{2020PASP..132j4401A} introduces a reddening-insensitive index, referred to as the $Q$ index, which serves as a proxy for the water absorption feature inherently present in young substellar objects. $Q$ is defined as

\begin{equation}
  Q= J-W + e(H-W),
\end{equation}

where $J$, $H$, and $W$ represent photometric data in their respective bands. The coefficient $e$ is defined as
\begin{equation}
  e= (A_J-A_W)/( A_W-A_H).
\end{equation}
$A_\lambda$, the selective extinction in each filter ($J$, $H$, and $W$)   derived  by \citet{2020PASP..132j4401A} using the spectrum of a M dwarf, its flux  unreddened and reddened by  $A_V =10$~mag adopting the  reddening law ($R_V =3.1$) from \citet{1999PASP..111...63F}. We use their value of $e = 1.85$.
The error in the $Q$ index is derived by propagating photometric errors from the $J$, $H$, and $W$ bands.

\begin{equation}
  \sigma_Q= \sqrt{ (\sigma_J^{2}+(e\sigma_H)^{2} +(1+e)\sigma_W^{2})}.
\end{equation}

\begin{figure}[htb!]
\includegraphics[width=0.52\textwidth]{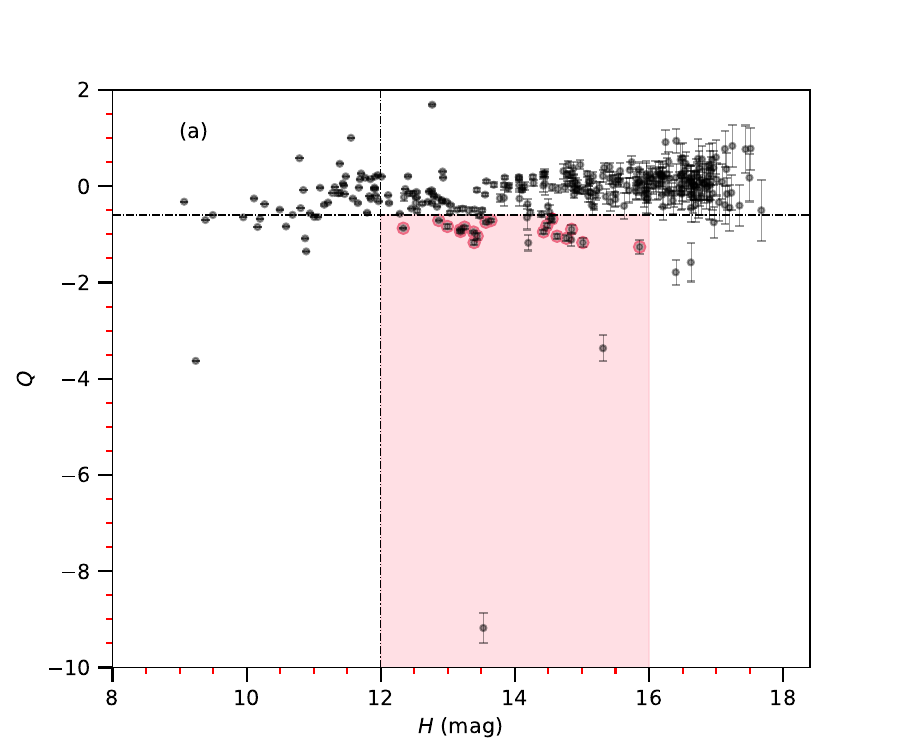}
\includegraphics[width=0.43\textwidth]{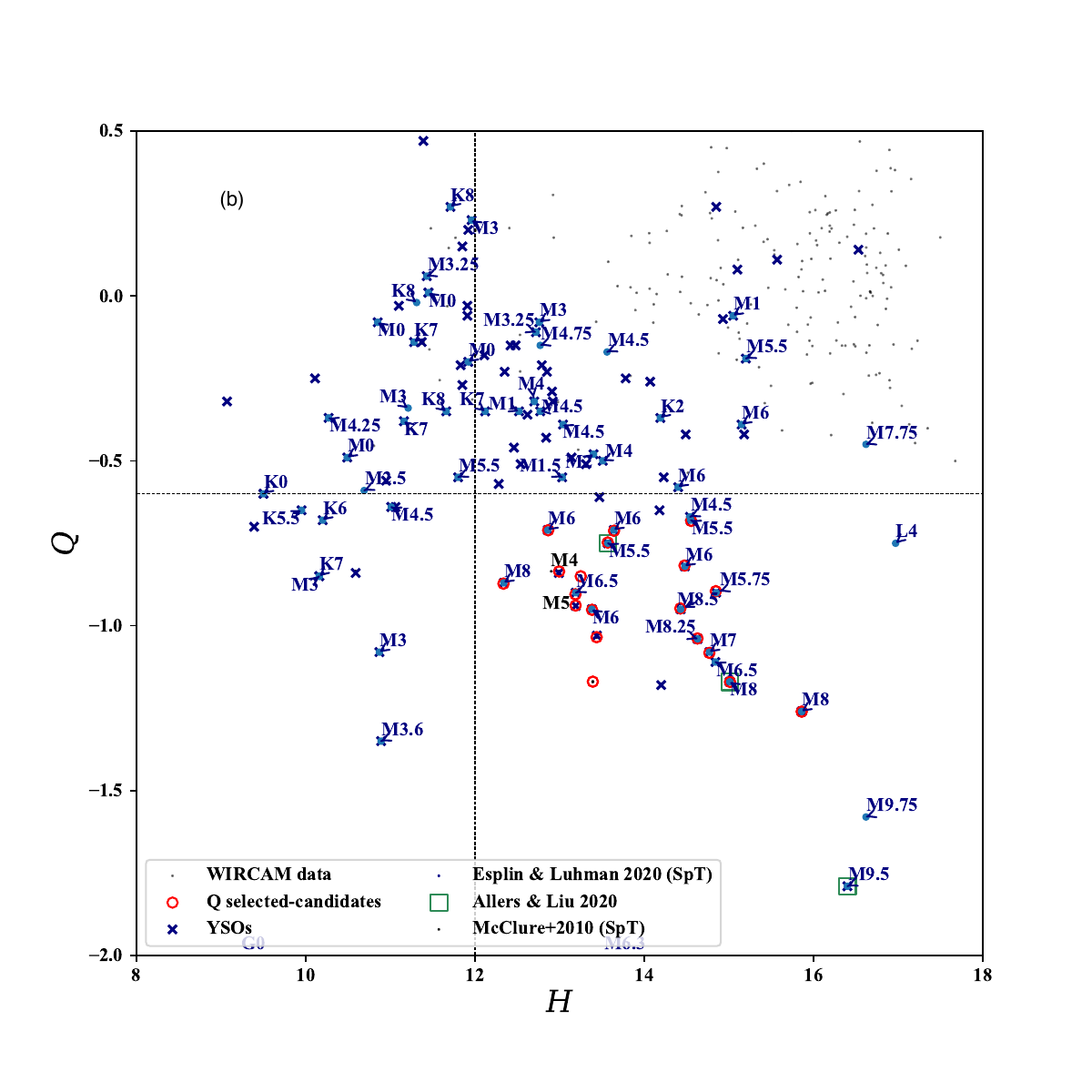}
\caption{(a)~The water $Q$ index versus $H$~mag for all WIRCam sources (in black). The $Q$ candidates are marked in red.  The pink shaded region highlights the region $Q< 0.6$~mag and $H$~mag between 12 and 16.  (b)~The zoomed-in region around our candidates where spectral types in the literature if available are labeled. }
    \label{Q_vs_H}
\end{figure}

With no water absorption present in their spectra, typical field M dwarfs are expected to have $Q\approx0$, and the value  progressively decreasing to $-0.6$ for M6 type, and  $-1$ for L0 types.  Substellar candidates therefore are selected by (1)~$Q < -(0.6+\sigma_Q$), i.e., for the water feature including the error, (2)~$H\ga12$~mag to avoid saturation, and (3)~$H \la 16$~mag for reliable photometry; equivalently this corresponds to about an error $\la 0.1$~mag in each band. The 19 $Q$ candidates thus identified (Figure~\ref{Q_vs_H}(a))  possess a $Q$ value between -0.6 and -1.8. There are two sources  with a  high negative $Q$ values (-3.5 and -9.3), and are not selected as per the criteria.  Figure~\ref{Q_vs_H}(b) labels sources with literature spectral types, namely by \citet{2010ApJS..188...75M, 2020AJ....159..282E, 2020PASP..132j4401A}, including 16 of our candidates.   In addition, there are three sources, though fainter than $H=16$~mag, included in our sample because they have $Q< 0.6$ with known spectral types of late M to L4.  Figure~\ref{Q_vs_H} exhibits a  general trend of $Q$ varying with the spectral type and with the $H$~mag for the low-mass objects and BDs in the cloud.  Consequently, our final list consists of a total of 22 young BD candidates, whose coordinates, $H$ band magnitude, $Q$ values, and SIMBAD identifiers, are presented in Table~\ref{table_Qcandidates}, ordered by the spectral type.


\begin{longrotatetable}
\begin{deluxetable}{cLLccCCl}
\tabletypesize{\scriptsize}
\tablecaption{Water bearing BD candidates}\label{table_Qcandidates}
\tablenum{1}
\tablehead{
\colhead{No.} & \colhead{$\alpha$~ (J2000)} & \colhead{$\delta$ ~ (J2000)}  & \colhead{SpTy} & \colhead{ref SpTy  \tablenotemark{$^\dagger$}} & \colhead{$H$}  & \colhead{$Q$} & \colhead{Simbad identifier}   \\
\colhead{} & \colhead{ } & \colhead{} & \colhead{} & \colhead{ } & \colhead{ ~ (mag)}   & \colhead{ } & \colhead{}}
\startdata
1 & 246.76716 & -24.47522 & M4 & 1 & 13.08~(0.03) & -0.835~(0.049) & SSTc2d J162704.1-242830 \\
2 & 246.60537 & -24.41251 & M5 & 1 & 13.10~(0.04) & -0.938~(0.037) & [GY92] 29 \\
3 & 246.65755 & -24.65098 & M5.5 & 2 & 13.57~(0.04) & -0.747~(0.025) & SSTc2d J162637.8-243903 \\
4 & 246.59925 & -24.30808 & M5.5 & 2 & 14.58~(0.06) & -0.682~(0.062) & SSTc2d J162623.8-241829 \\
5 & 246.67014 & -24.51427 & M5.75 & 2 & 14.87~(0.06) & -0.895~(0.086)& SSTc2d J162640.8-243051 \\
6 & 246.59567 & -24.47951 & M6 & 2 & 12.80~(0.02) & -1.082~(0.060) & SSTc2d J162622.3-242407 \\
7 & 246.59248 & -24.39787 & M6 & 2 & 13.50~(0.03) & -0.712~(0.027) & [GY92] 10 \\
8 & 246.57909 & -24.40408 & M6 & 2 & 13.75~(0.03) & -0.951~(0.025) & SSTc2d J162619.0-242414 \\
9 & 246.63552 & -24.44321 & M6 & 2 & 14.46~(0.05) & -0.817~(0.052) & [GY92] 64  \\
10 & 246.57842 & -24.43637 & M6.5 & 2 & 13.20~(0.03) & -0.904~(0.016) & SSTc2d J162618.8-242610 \\
11 & 246.59275 & -24.40192 & M7 & 2 & 14.83~(0.01) & -1.082~(0.06) &  SSTc2d J162622.3-242407 \\
12 & 246.86070 & -24.43190 & M8 & 2 & 12.35~(0.02) & -0.872~(0.005) & SSTc2d J162726.6-242554 \\
13 & 246.57738 & -24.49771 & M8 & 2 & 15.07~(0.01) & -1.170~(0.096) & SSTc2d J162618.6-242951 \\
14 & 246.77472 & -24.31119 & M8 & 2 & 15.86~(0.02) & -1.260~(0.138)  & SONYC RhoOph-6 \\
15 & 246.71371 & -24.54503 & M8.5 & 2 & 14.47~(0.05) & -0.947~(0.044) & [RR90] Oph 2349.8-2601 \\
16 & 246.51367 & -24.50717 & M9.5 & 2 & 16.45~(0.03) & -1.790~(0.260) & SONYC RhoOph-7 \\
17 & 246.66634 & -24.37600 & M8.25 & 2 & 14.63~(0.04) & -1.038~(0.050) &  [GY92] 90  \\
18 & 246.85683 & -24.62461 & M9.75 & 2 & 16.60~(0.03) & -1.580~(0.400) & CFHTWIR-Oph  77 \\
19 & 246.66536 & -24.36837 & L4 & 2 & 16.94~(0.04) & -0.750~(0.320) & CFHTWIR-Oph  33 \\
20 & 246.77487 & -24.43850 & \nodata & \nodata & 13.34~(0.03) & -1.034~(0.072) & YLW10A \\
21 & 246.70307 & -24.63997 & \nodata & \nodata & 13.46~(0.00) & -0.850~(0.028) & BBRCG 8 \\
22 & 246.62528 & -24.64658 & \nodata & \nodata & 13.88~(0.20) & -1.169~(0.049) & \nodata \\
\enddata
\label{Tab:all_list}
\tablenotetext{$$^\dagger$$}{The reference for spectral types 1: \citet{2010ApJS..188...75M}, 2: \citet{2020AJ....159..282E} }
\end{deluxetable}
\end{longrotatetable}

\subsection{Multiepoch Astrometry and Motions} \label{sec:astrometric data}

We further ascertained the membership of our young BD candidates by their motion.  L\,1688A has significant extinction, 50~mag or more in certain parts of the cloud (Figure~\ref{fov_data}), rendering Gaia data unavailable for the majority of our candidates.  Lacking spectroscopically confirmed AGNs or QSOs in our field, we derived astrometry and hence proper motion of faint sources by using the CADC archives of deep WIRCam Ks-band images spanning six years from 2006 to 2012. The images for each epoch were stacked to generate a catalog of point sources using Sextractor \citep{1996A&AS..117..393B}.  Gaia DR3 \citep{2022yCat.1355....0G} astrometry was exercised for calibration of bright sources to derive the motion of faint candidates, described below.  

\begin{figure*}[htb!]
    \includegraphics[width=0.5\textwidth]{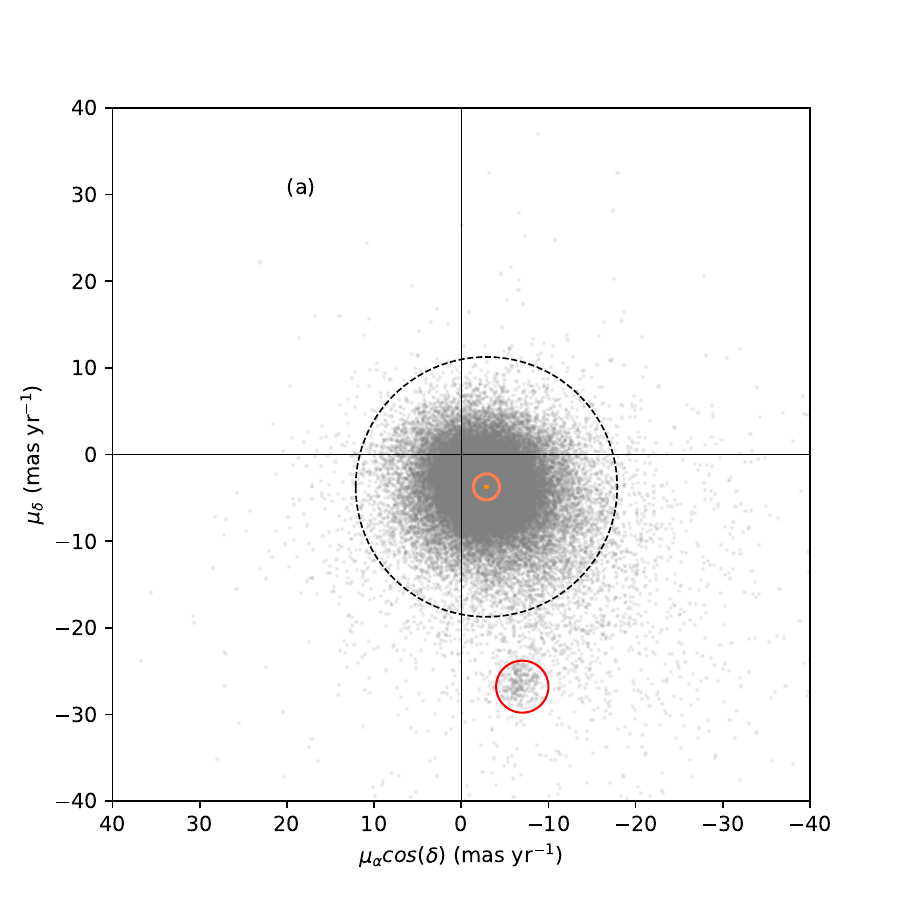} %
  \includegraphics[width=0.5\textwidth]{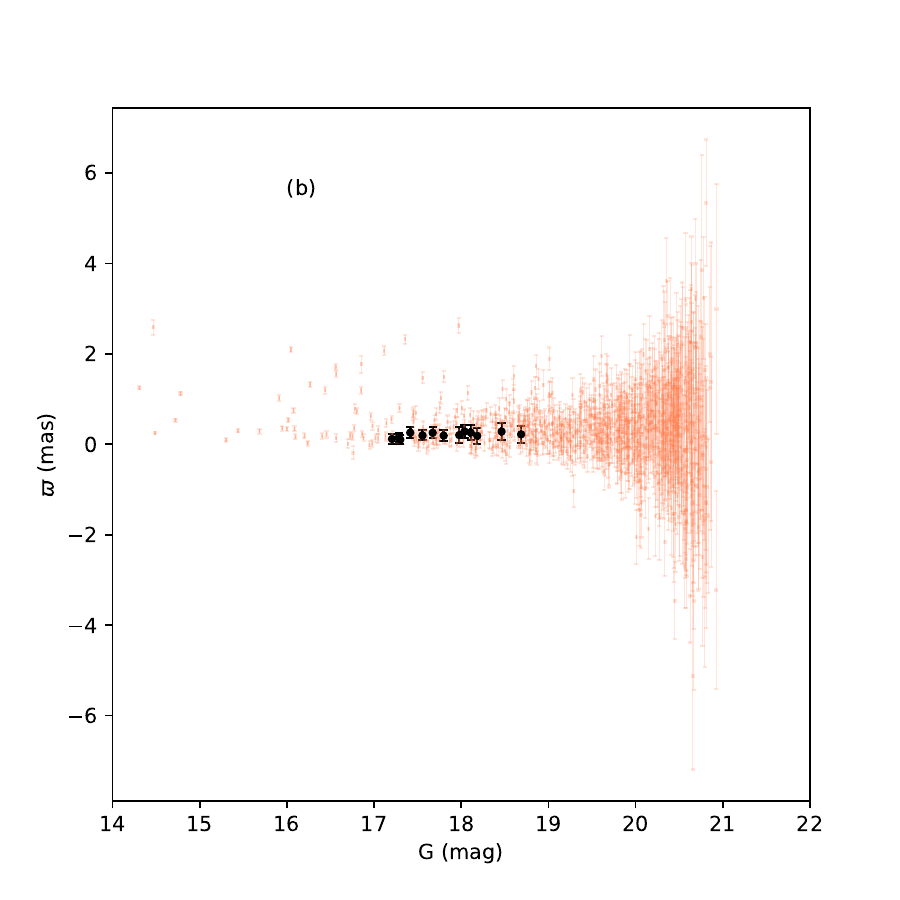}%
    \caption{(a)~Proper motion vector plot for Gaia DR3 sources. In addition to the L\,1688A cluster in the lower right quadrant (red circle), the field population is concentrated.  Stars inside the black dashed circle are used to compute a centroid sample, marked in orange, to represent ``motionless'' objects.  (b)~The parallax versus $G$ band magnitude of the motionless objects selected in (a).  Among these, 13 bright and distant objects, marked in black, are considered high-confidence stationary objects as proper motion standards.}
    \label{sel_motionless}
\end{figure*}

First, a group of ``stationary'' field objects were selected based on the Gaia proper motion. Figure~\ref{sel_motionless}(a) illustrates the segregation of cluster members from field stars.  The field sample does not center at the origin as a result of the solar reflex motion.  Near the center of this concentration at ($\mu_{\alpha} \cos\delta, \mu_{\delta}) = (-2.8, -3.7)$~mas~yr$^{-1}$ are stars with null proper motion relative to the Sun.  Next, within a radius of 1.5~mas~yr$^{-1}$ of the center, we selected relatively bright Gaia stars (for which measurements are reliable) with nearly negligible parallax values and minimal errors (distant thereby with little expected proper motion).  A total of 13 such high-confidence stationary Gaia objects are used to calibrate the astrometry with which to derive the proper motion in the deep infrared images.  

We measured the device coordinates of all infrared sources relative to bright Gaia motionless standards, to derive proper motions between two epochs six years apart.  The final value is the average of all 13 stationary objects, multiplied by the plate scale ($0\farcs3$), leading to proper motions in milliarcseconds per year (mas~yr$^{-1}$).

\begin{figure}[htb!]
    \includegraphics[width=0.58\columnwidth]{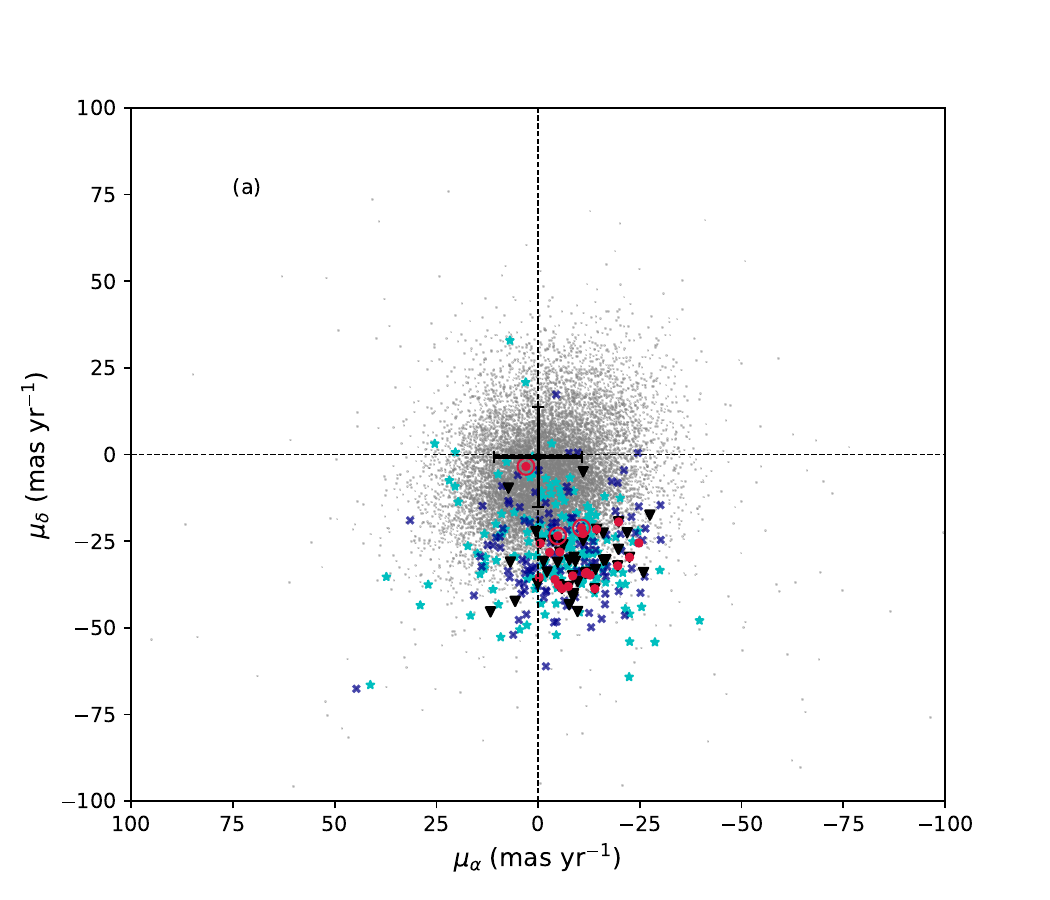}    \includegraphics[width=0.42\columnwidth]{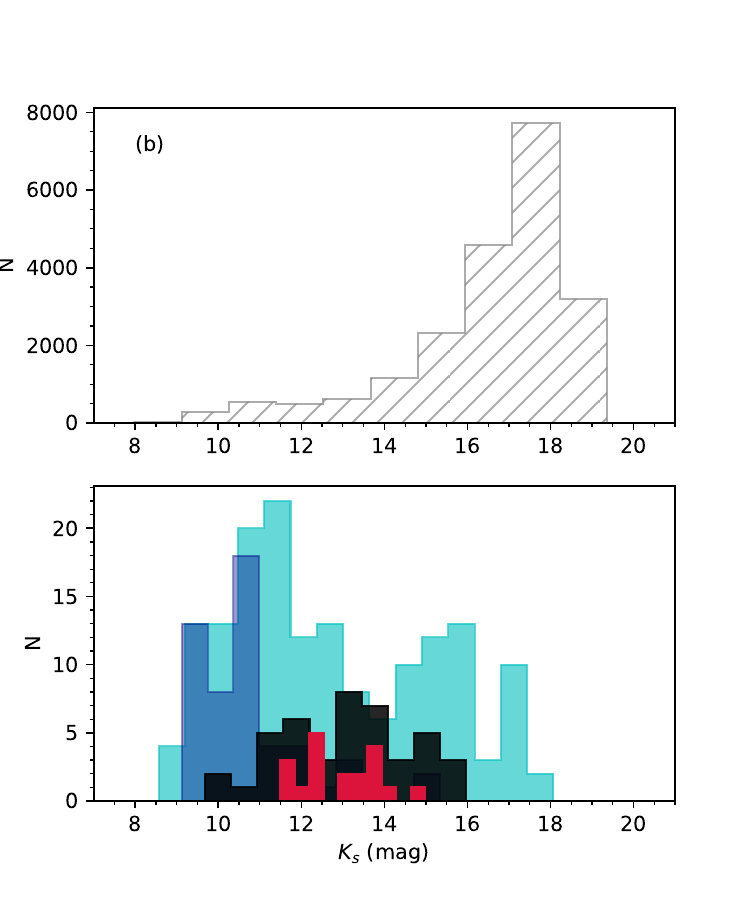} 
    \caption{(a)~All infrared sources (grey points) for which proper motions have been computed in this work, including our 22 BD candidates in red.  Also marked are the 115 \cite[Gaia, in blue]{2021A&A...652A...2G} and 43 kinematic samples of YSOs \cite[Spitzer, in black]{2020AJ....159..282E}.  The 161 candidates/members without previously reported proper motion are represented in cyan. The average of the motionless (field) sources is near $(0,0)$ along with the corresponding dispersion (indicated in black). (b)~The $K_s$~mag distribution for, in the top panel, all infrared sources (grey), and, in the bottom panel, for young samples with the same color scheme as in (a)).}
    \label{computed_motion}
\end{figure}

Figure~\ref{computed_motion}(a) exhibits the proper motion of sources derived from our images, including the YSOs compiled by \citet{2020AJ....159..282E} using motion from Spitzer's multiepoch observations, and those by \citet{2021A&A...652A...2G} using Gaia data of known members to identify a large sample of new candidates.  In this revised vector point diagram of proper motion, the data points are concentrated near the origin, denoting the field population, less the solar motion. In contrast, the motion of the YSOs with known motion clearly stands out, averaging ($\mu_{\alpha} \cos\delta, \mu_{\delta}) = (-6.7, -28.7)$~mas~yr$^{-1}$ with standard deviations of ($\mu_{\alpha} \cos\delta, \mu_{\delta}) = (11.6, 12.4)$~mas~yr$^{-1}$. Our 22 BD candidates, 3 of which have Gaia data and 13 with Spitzer measurements, have consistent kinematic distribution with those of the YSO samples. The kinematic membership of six (not in Gaia or Spitzer) are established in this work, five of which (sources No.~1, 2, 19, 20, 21) are associated with cluster, whereas (No.~22) is not. Figure~\ref{computed_motion}(b) indicates that while our BD sample is fainter than Gaia, it still  overlaps with brightness range covered by Spitzer's. However, this analysis allowed us to derive, the proper motion of additional 161 young fainter candidates without prior information on kinematics from Gaia or Spitzer.

\subsection{Multiwavelength Archival Data and Dusty Disks} \label{sec:multi_wav data}

\begin{figure*}[htb!]
    \includegraphics[width=.32\textwidth,height=3.3cm]{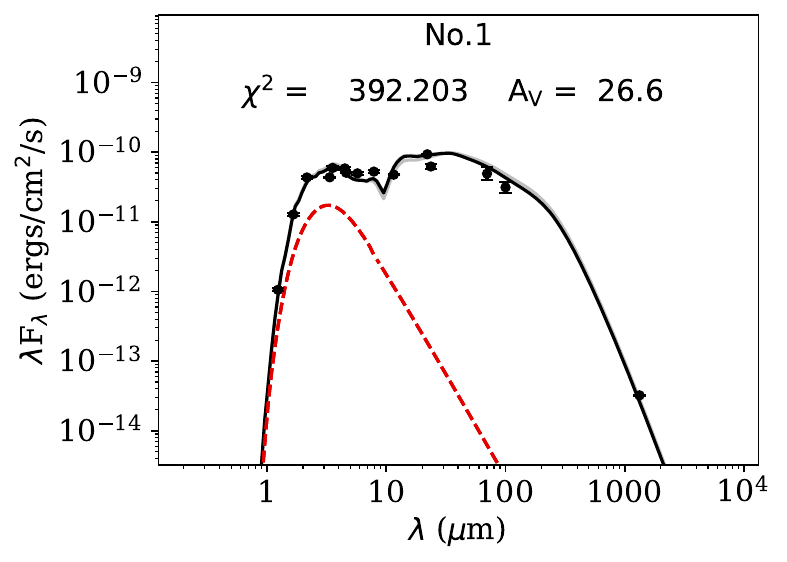}
    \includegraphics[width=.32\textwidth,height=3.3cm]{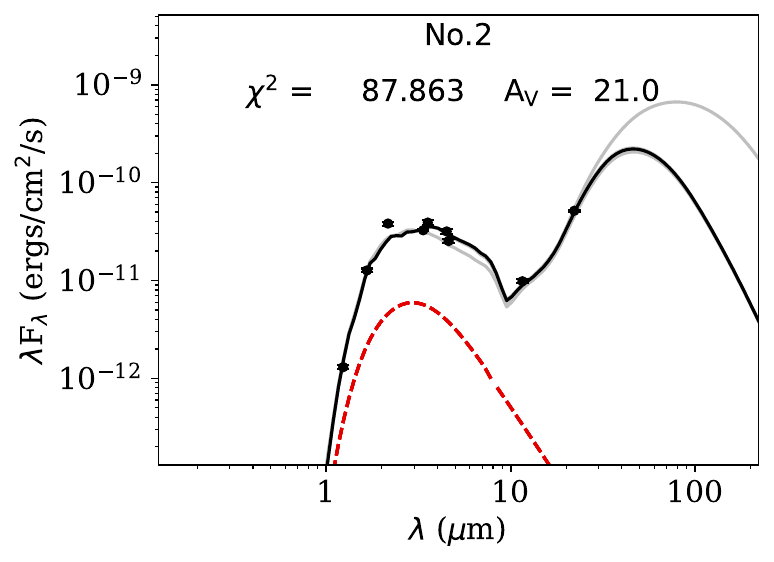}
    \includegraphics[width=.32\textwidth,height=3.3cm]{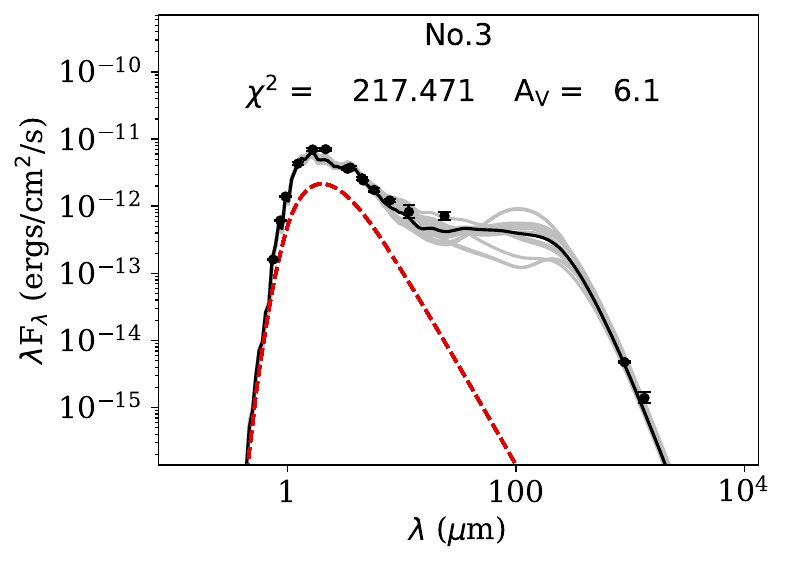}
    \includegraphics[width=.32\textwidth,height=3.3cm]{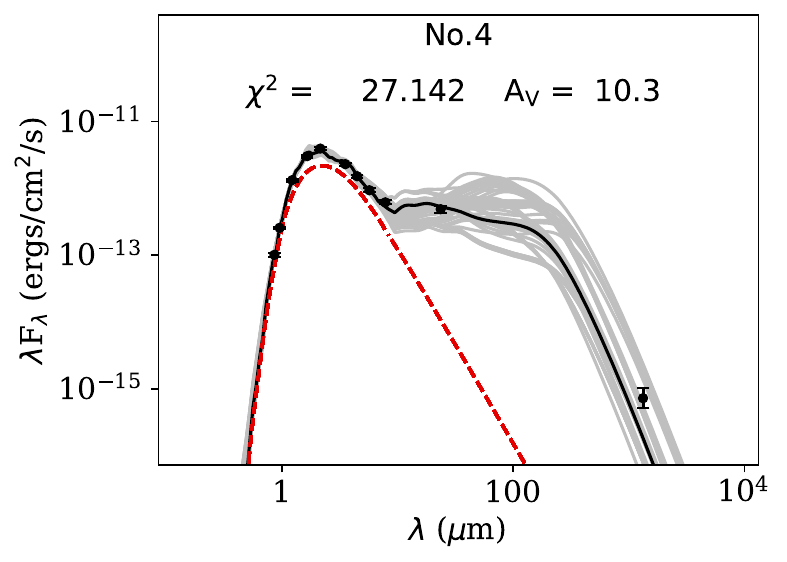}
    \includegraphics[width=.32\textwidth,height=3.3cm]{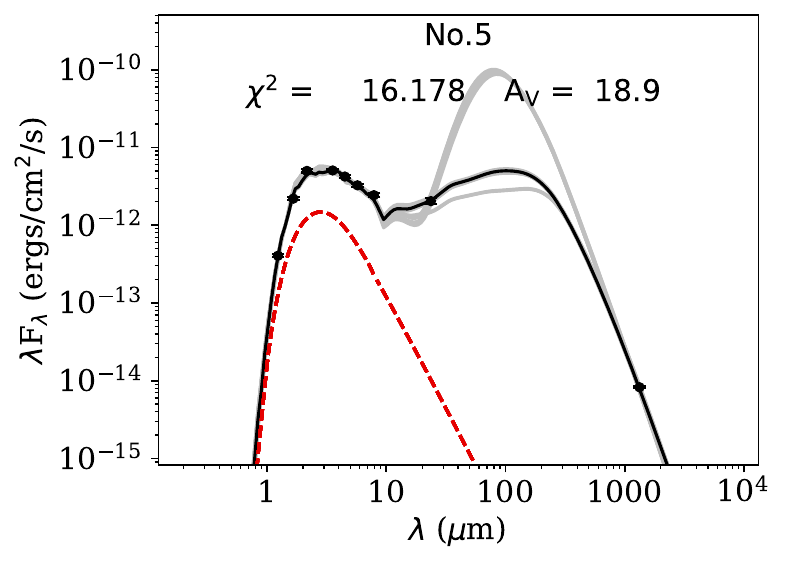}
    \includegraphics[width=.32\textwidth,height=3.3cm]{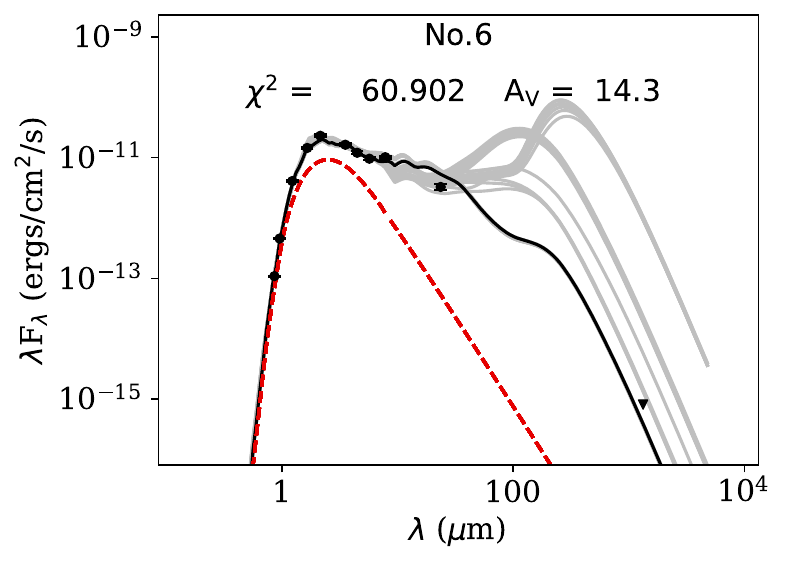}    \includegraphics[width=.32\textwidth,height=3.3cm]{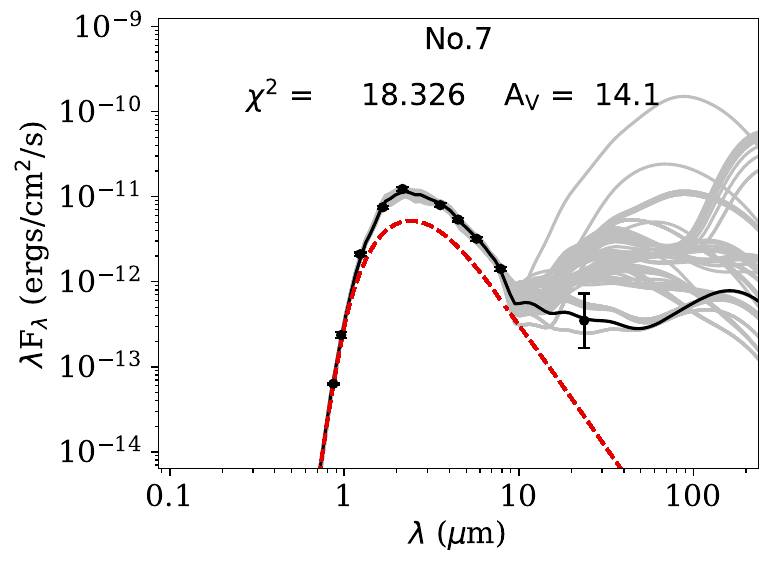}
    \includegraphics[width=.32\textwidth,height=3.3cm]{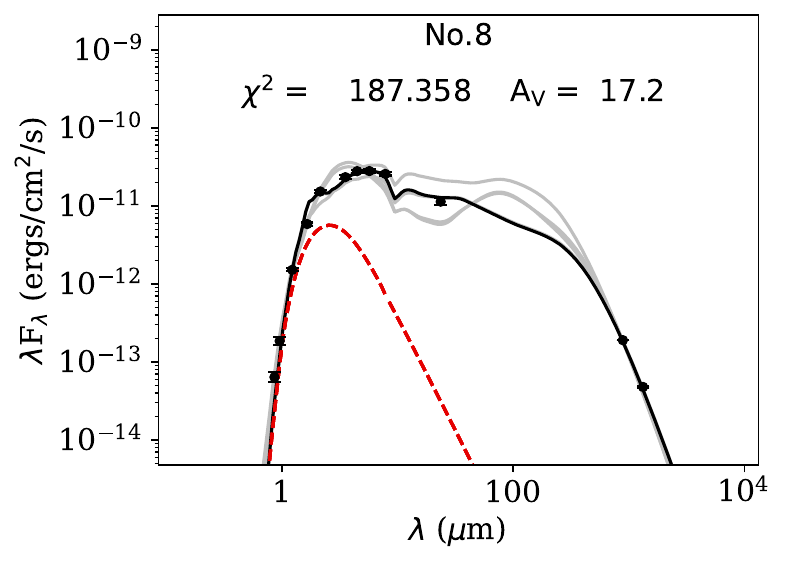}
    \includegraphics[width=.32\textwidth,height=3.3cm]{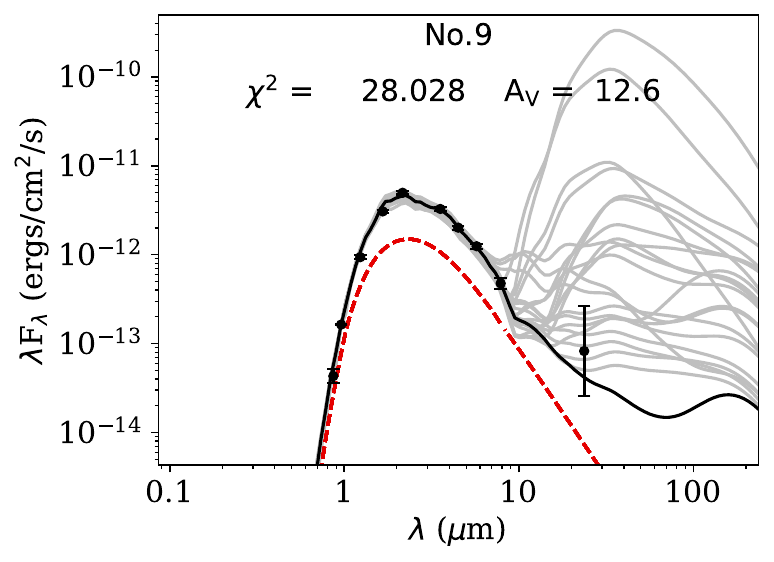}
    \includegraphics[width=.32\textwidth,height=3.3cm]{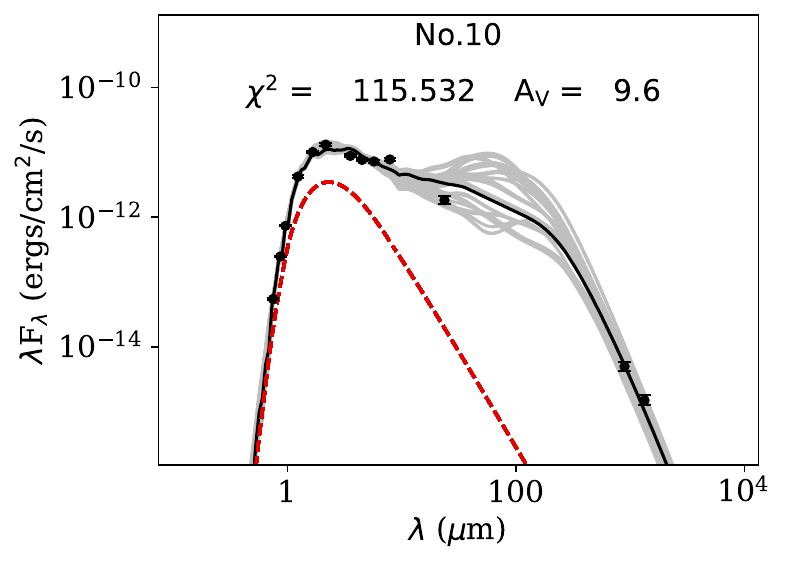}
    \includegraphics[width=.32\textwidth,height=3.3cm]{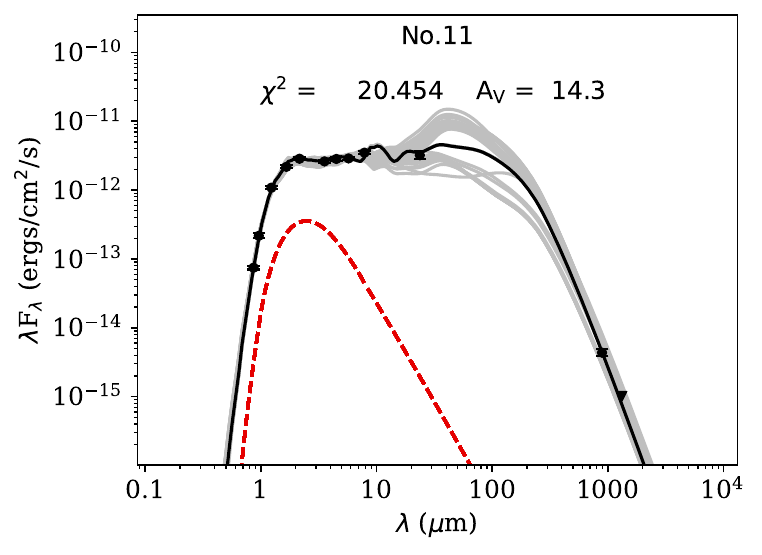}
    \includegraphics[width=.32\textwidth,height=3.3cm]{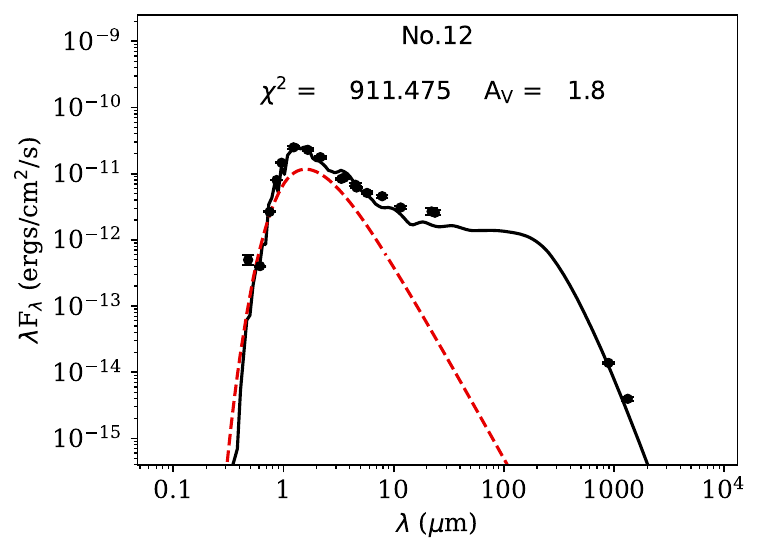}    \includegraphics[width=.32\textwidth,height=3.3cm]{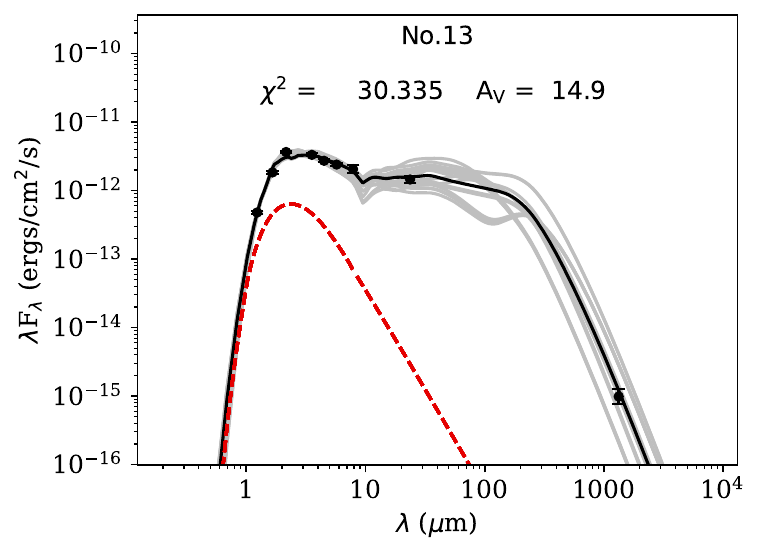}
    \includegraphics[width=.32\textwidth,height=3.3cm]{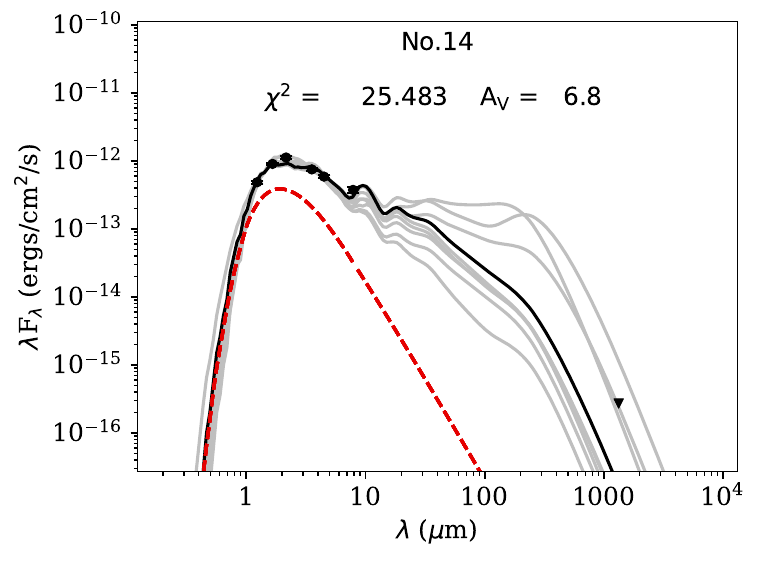}
    \includegraphics[width=.32\textwidth,height=3.3cm]{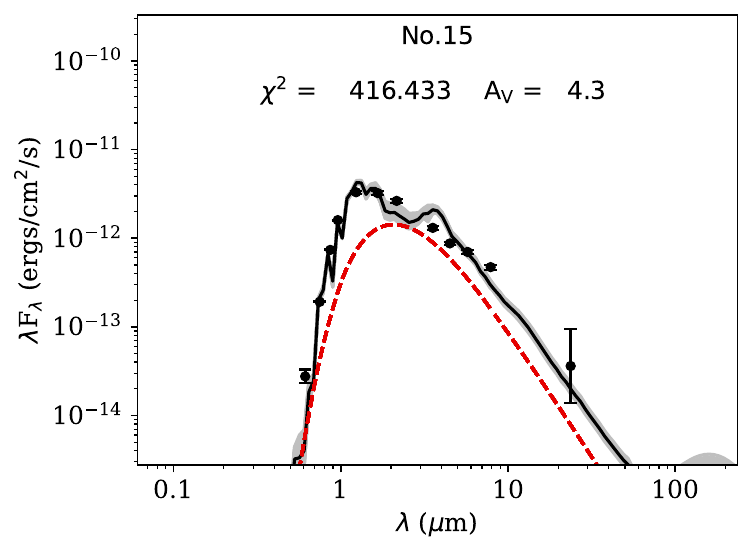}
    \includegraphics[width=.32\textwidth,height=3.3cm]{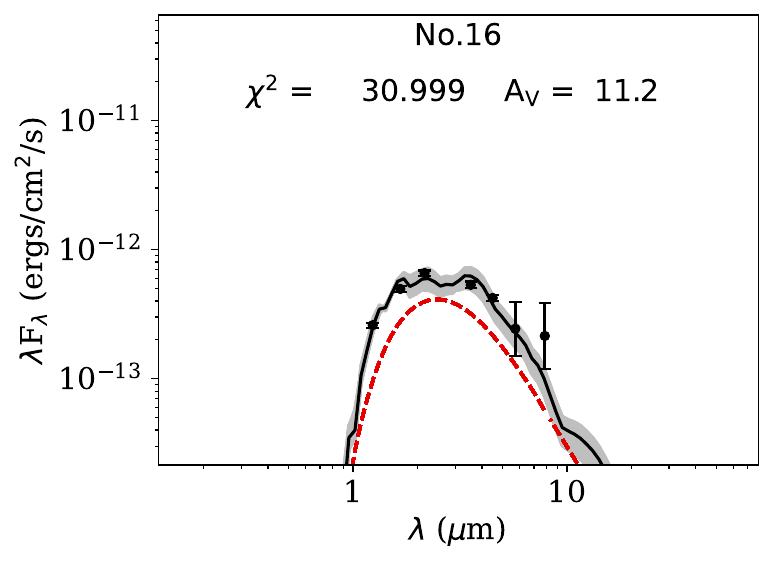}
    \includegraphics[width=.32\textwidth,height=3.3cm]{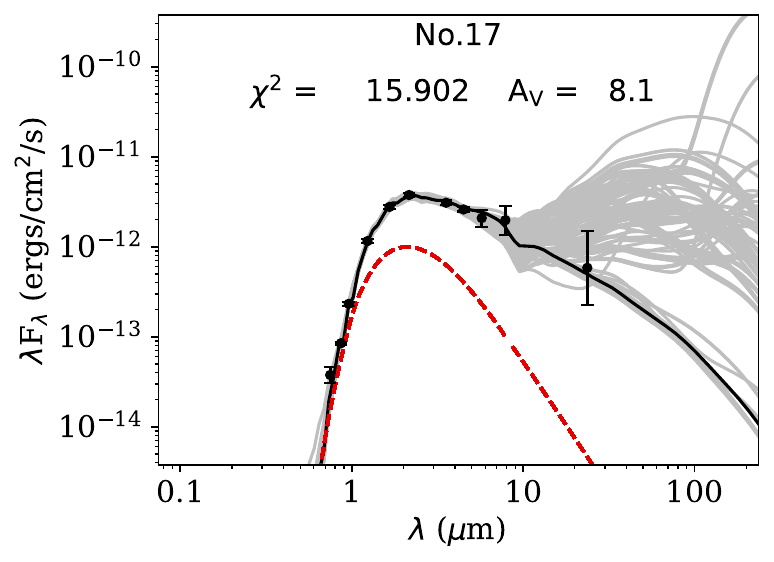}
    \includegraphics[width=.32\textwidth,height=3.3cm]{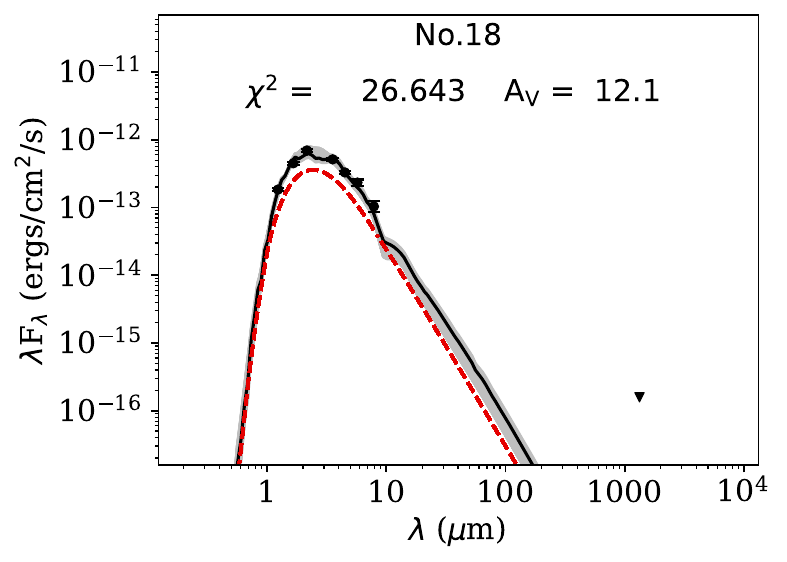}  
    \includegraphics[width=.32\textwidth,height=3.3cm]{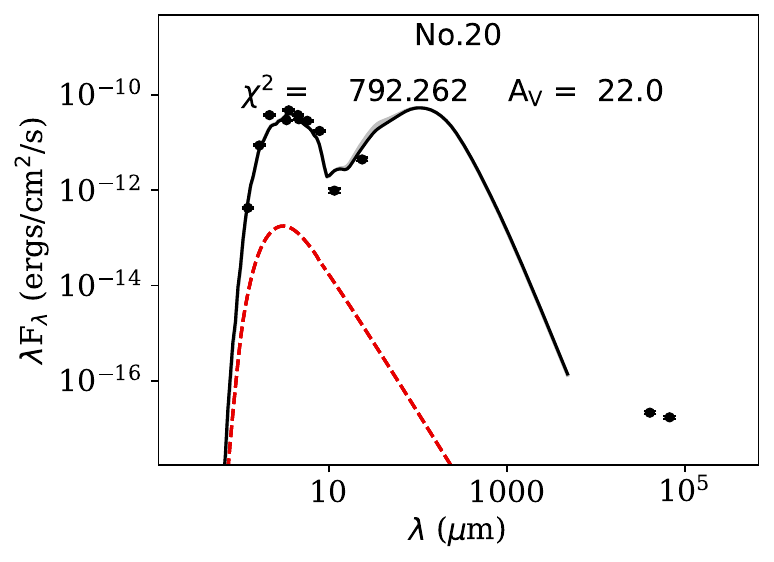}   
    \caption{SEDs of all objects with the internal id from Table~\ref{table_Qcandidates}, and best fit parameters on each figure. The observed data points (N) are shown in black with best fit model curve in black, and in fairly good fit models with $\chi^{2}- \chi_{min}^{2} <3N$ is indicated in grey. The red dashed line represent the best fit blackbody curve. }\label{SEDS}
\end{figure*}

To diagnose possible existence of dusty disks in the substellar objects, we have collected photometry for each source using Vizier photometry viewer \citep{2014ASPC..485..219A} that includes  optical from Pan-STARRS \citep{2016arXiv161205560C}, near-infrared from Two Micron All Sky Survey (2MASS) \citep{2003yCat.2246....0C}, mid-infrared from Spitzer \citep{2014yCat..51480122G} and WISE \citep{2012yCat.2311....0C}, far-infrared from Herschel \citep{2020yCat.8106....0H}, and ALMA 1.3~mm fluxes \citep{2019MNRAS.482..698C} when available.  The SEDs thus obtained is fit for each source with YSO models from \citet{2017A&A...600A..11R}, which for best results requires data points to have a minimum of five data points and sufficiently wide wavelength coverage, particularly in near-infrared and mid-infrared ranges which is available for 19 of our sources.  The fitter treats both distance and extinction as free parameters.  In our analysis we adopted a distance range of 120~pc to 150~pc, and an extinction between 0 and 40~mag.  
 The SED and the best-fit model for each of the  sources are presented in Figure~\ref{SEDS}.  15 $Q$ selected objects show evidence of disks based on their SEDs.

\begin{figure}[htb!]
    \includegraphics[width=\columnwidth]{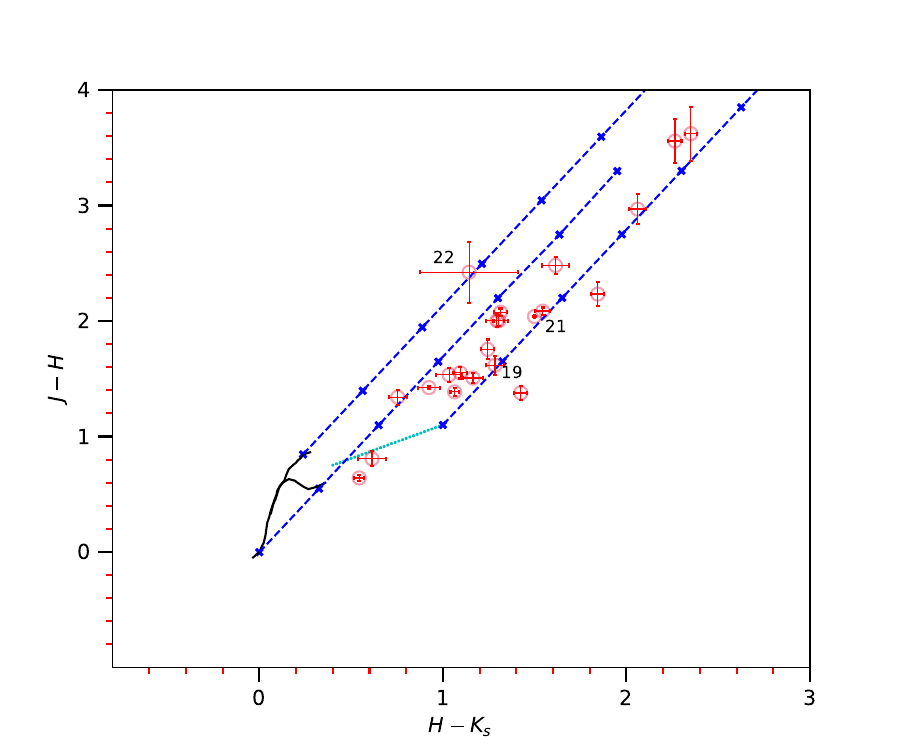} %
    \caption{Near-infrared 2MASS $J-H$ versus $H-K_s$ color-color diagram. All of our $Q$ candidates are marked by red circles, with the three sources not analyzed by the SEDs are denoted by black squares, and are individually labeled. For reference, the black curves represent unreddened main sequence and giant star loci \citep{1988PASP..100.1134B}, whereas the intrinsic locus of classical T~Tauri stars is represented by the cyan line \citep{1997AJ....114..288M}. The dashed blue lines indicate the direction of reddened objects along the reddening vector, adopting an interstellar law \citep{1981ApJ...249..481C}, with blue crosses marked  at intervals corresponding to $A_V$ of 5~mag.    
    }
\label{nir_tcd}
\end{figure}

Three sources (No.~19, 21, 22) lack data of sufficient wavelength coverage to render reliable SED fitting.  For these, we inspected the color-color diagram (Figure~\ref{nir_tcd}), which serves as a simplified tool, though less robust than SEDs, to estimate possible infrared excesses originating from inner disks. 
Our color analyses reveals that one of the three candidates (i.e., No.~22) is devoid of infrared excess, appears to be a  reddened giant star. No.~22 also exhibits an  inconsistent proper motion, making it a contaminant in our BD sample.  Overall, out of 22 candidates, 15 manifest dusty disks clearly evidenced by SEDs. Six sources exhibit near-infrared excess  but lacks sufficient data at long wavelengths to be confirmed by SEDs leaving them to be disk candidates.

\begin{figure}[htb!]
    \includegraphics[width=\columnwidth]{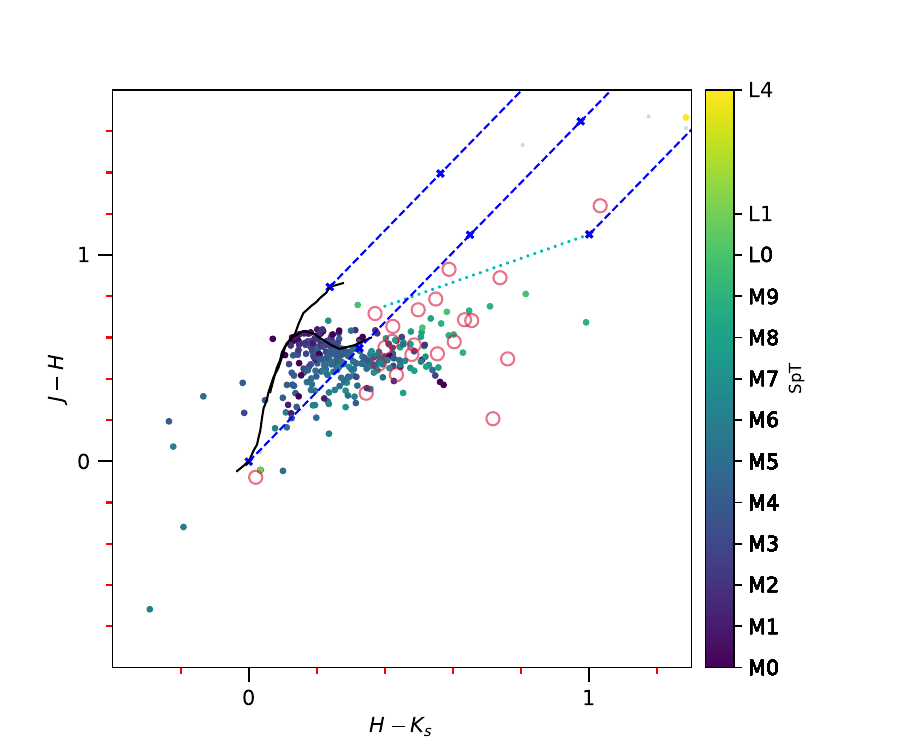} %
\caption{The same as in Figure~\ref{nir_tcd} with $Q$ candidates (red circles) and M dwarfs dereddened and color coded with respective spectral types. 
    }
\label{nir_tcd_dered}
\end{figure}

We  also present a comparison of  intrinsic colors of our BD sample with the dwarf locus in Figure~\ref{nir_tcd_dered}. The dereddened colors for our sample are derived using extinction estimates from the SED fitting, and adopting the reddening law from  \citet{1981ApJ...249..481C}. The extinction values  and spectral type for M type  members of the complex are taken from \citet{2020AJ....159..282E}, which upon dereddening reproduces the dwarf locus of \citet{1988PASP..100.1134B} with an extension into classical T-Tauri region with near infrared excess. There is a continuous  trend in the spectral type on locus of M dwarf down to the substellar regime. The alignment of $Q$ candidates with the loci also suggests extinction from the SED fitting are reliable.

\subsection{Spectroscopic Observations} \label{sec:spectroscopy obs}

We also acquired medium-resolution IR spectra of eight very low-mass and substellar objects using the SpeX instrument \citep{2003PASP..115..362R} at Infrared Telescope Facility (IRTF). The observations were performed on 2023 August 3 and August 7. The short wavelength ($0.25-2.5~\micron$) coverage in cross-dispersed mode with 0.8\arcsec slit was utilized to achieve a spectral resolution of 750. The $K_s$~mag range of the targets was between 11~mag and 13.5~mag, with the integration time ranging from 1080~s to 1800~s. For telluric corrections, an A0V star (HD\,147384) was observed throughout the observing run with an airmass $<0.1$ and an angular separation $<5$~deg from each science target. The data were reduced in the standard manner employing the reduction tool SpeXtool version 4.1 \citep{2004PASP..116..362C, 2003PASP..115..389V}. The observing log is presented in Table~\ref{table:IRTF_log}.

\begin{deluxetable}{cCCcCCl}
\tabletypesize{\scriptsize}
\tablecaption{IRTF Observing log}
\tablenum{2}
\tablehead{
\colhead{No.\tablenotemark{$^a$}}  &  \colhead{$\alpha$~(J2000)} & \colhead{$\delta$~(J2000)}  &\colhead{SpT \tablenotemark{$^b$}}   & \colhead{$K_s$} & \colhead{i time} & \colhead{Simbad identifier} \\
\colhead{}  & \colhead{} & \colhead{ } & \colhead{ }   &\colhead{(mag)}  &\colhead{(sec)} &\colhead{} }
\startdata
8 & 246.63552 & -24.44321& M6  &  11.9 & 1440 & [GY92] 64 \\
13 & 246.57738 & -24.49771 & M8  & 13.5  & 1200 & BKLT J162848 242631\\
17 & 246.66634 & -24.37600 & M8.25  & 13.5 & 1800 & [GY92] 90\\
20 & 246.77487 & -24.43850  & \nodata  & 11.1 & 1080 & YLW10A\\
\hline
23 & 247.05300 & -24.19322 & M6.25  & 11.1  & 1080 & ISO Oph 193 \\
24 & 247.20296 & -24.44226 & M6.25  & 13.1 & 1440 & CFHTWIR Oph107 \\
25 & 246.92016 & -24.48366 & M7.5  & 13.1 & 1440 & [GY92] 320 \\
26 & 246.76676 & -24.04642 & M7.25  & 13.2  & 1800 & CFHTWIR Oph57 \\
\enddata
\label{table:IRTF_log}
\tablenotetext{$$^a$$}{Source No.~8, 13, 17, 20 (upper panel) are  water bearing candidates  from Table~\ref{table_Qcandidates}. The remaining four (lower panel) are outside the field of W-Band imaging.}
\tablenotetext{$$^b$$}{Spectral types from \citet{2020AJ....159..282E} }

\end{deluxetable}

\begin{figure}
    \centering
    {\includegraphics[width=\textwidth]{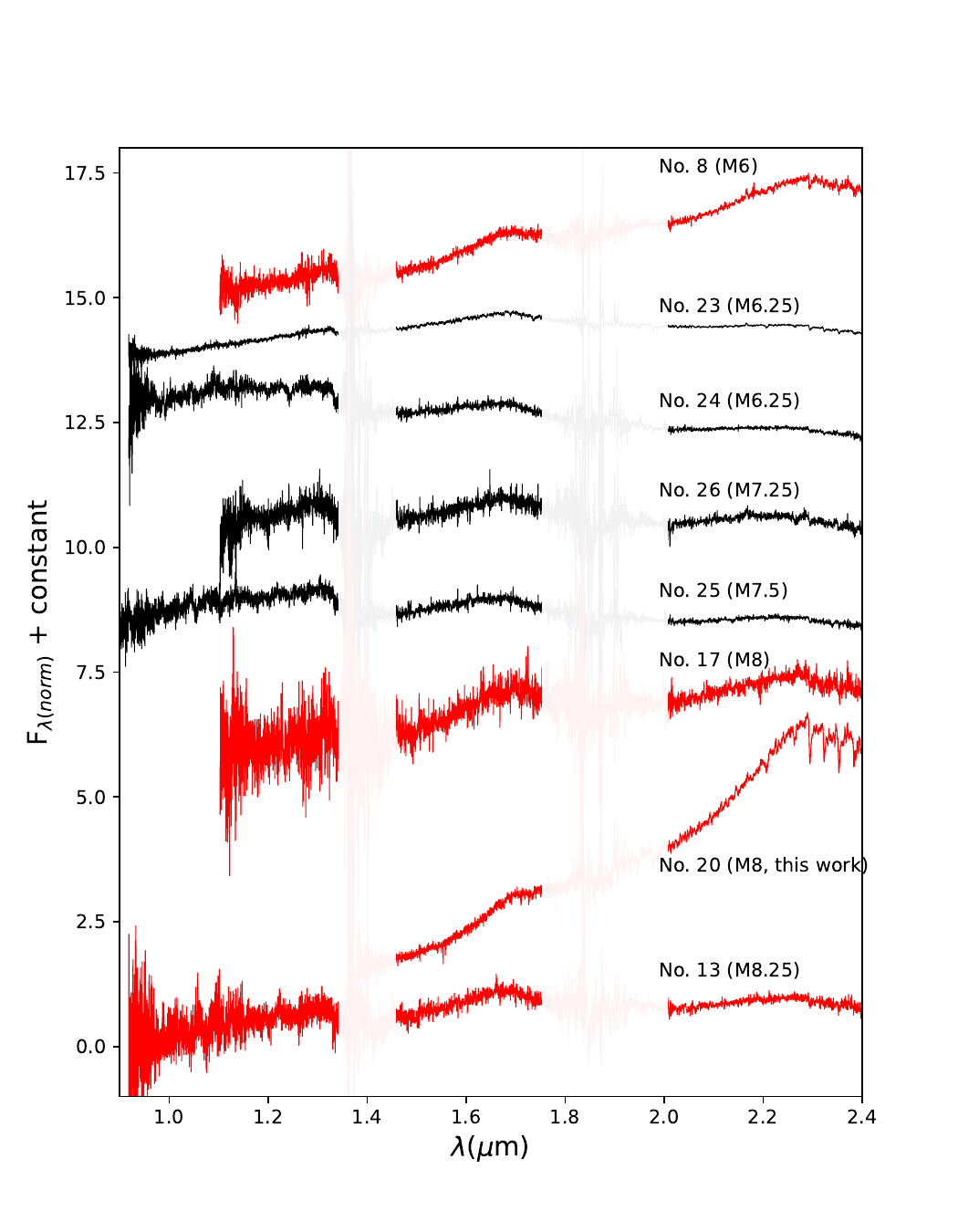}} 
    \caption{Reduced IRTF spectra of $Q$ candidates (in red) and known low-mass objects outside the W-band image (in black), each with the source ID (as per Table~\ref{table:IRTF_log}) and spectral type labeled.  No.~20 has poor data quality in the $J$ band, so only the long-wavelength spectrum is shown.
    }
    \label{reduced_spectra}
\end{figure}

Among the eight spectroscopic targets, four were $Q$ selected candidates, and four were known substellar objects but outside the field of the W-band image.  All but one (No.~20) have had spectral typing available in the literature.  The purpose of the spectroscopy, in addition to ascertain the substellar nature, is to detect possible accretion features, such as Pa$\beta$ and Br$\gamma$ emission lines. 
The reduced spectra are exhibited in Figure~\ref{reduced_spectra}, in order of the spectral type, with object No.~20 determined to be an M8.

\begin{figure*}[htb!]
  \includegraphics[width=0.45\textwidth]{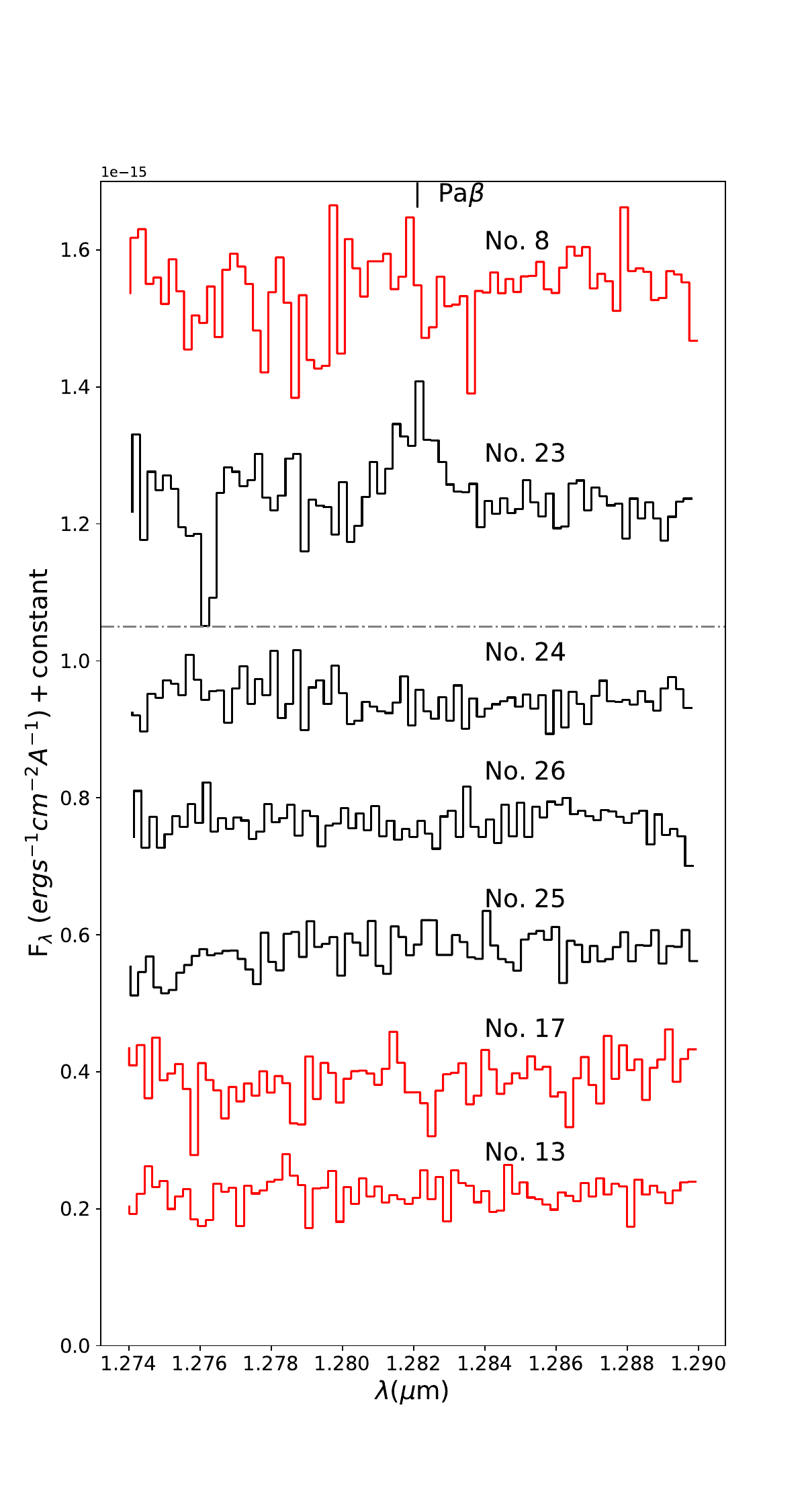}%
  \includegraphics[width=0.45\textwidth]{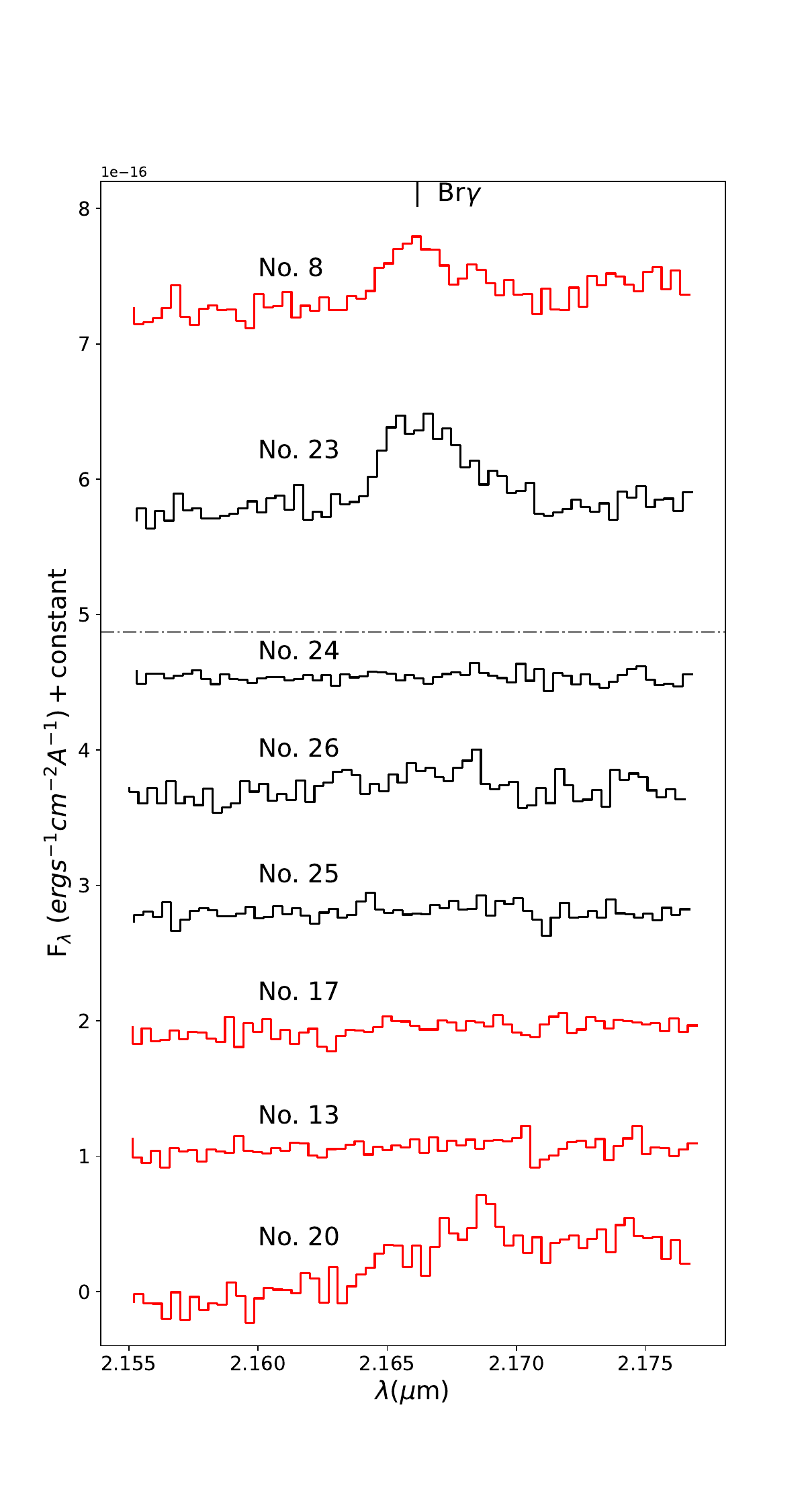}%
    \caption{The same spectra as in Figure~\ref{reduced_spectra} but zoomed in to near Pa$\beta$ on the left panel and Br$\gamma$ on the right panel (less No.~20). Accretion features are detected in the top two sources, namely, No.~8 and No.~23. }
    \label{all_tar_pab_}
\end{figure*}

The inverted-V shape near 1.67~\micron~in the spectra characteristic of young substellar nature  \citep[e.g.,][]{2001MNRAS.326..695L, 2004ApJ...617..565L,2006ApJ...639.1120K, 2007ApJ...657..511A} can be seen in most targets, and more prominent for later spectral types.  Figure~\ref{all_tar_pab_} zooms in near the Pa$\beta$ and Br$\gamma$ lines where magnetospheric accretion shocks would produce these lines in emission.  

Target No.~8  is known to have Pa$\beta$ \citep{2006A&A...452..245N},  while No.~23 is known to have both Pa$\beta$ and Br$\gamma$ emission \citep{2006A&A...452..245N, 2006A&A...460..547G}.
 In our data, No.~23 displays clear emission indeed in Pa$\beta$ (Figure~\ref{all_tar_pab_}), with an EW of $1.86 \pm 1.39$~\AA. However, in No.~8 Pa$\beta$ emission is not as clear, with the EW of $3.54 \pm 3.98$~\AA. Both these measured values are consistent with the literature.  Our spectra of both targets show clear evidence of accretion in Br$\gamma$. To our knowledge, our observations present the first Br$\gamma$ detection for No.~8. The equivalent width (EW) is determined by integrating the area under the line after subtracting the local continuum flux. Subsequently, the line flux ($f_{\rm line}$) is calculated by multiplying the EW with the extinction-corrected flux, using  $A_J$ and $J$ for Pa$\beta$, and $A_{K_s}$ and $K_s$ for Br$\gamma$, where $A_J$  and $A_{K_s}$ represent the extinction in $J$ and $K_s$ bands.  This line luminosity is then estimated by $4\pi d^2 f_{\rm line}$ assuming a distance of 137.3~pc  \citep{2017ApJ...834..141O}. The line luminosity is converted to accretion luminosity \citep{2004AJ....128.1294C, 2004A&A...424..603N}. For non-detections, the 3$\sigma$ upper limit on the EW is calculated as $3(\Delta f/f_{ctn}) \Delta \lambda$, where $\Delta f$ is the rms flux noise of each spectrum, $f_{ctn}$ is the continuum flux, and adopting $\Delta \lambda$ equal to the linewidth of 40~\AA. The corresponding accretion luminosity is calculated using the same methodology described for detected lines. The accretion luminosities from this work, and collected from literature \citep{2006A&A...452..245N, 2015A&A...579A..66M, 2024A&A...685A.118A} are listed in Table~\ref{table:Q_param}.

Table~\ref{table:Q_param} summarizes some properties of our $Q$ candidates.  A sequence number is assigned in Column~1, followed by the spectral type in Column~2. Column~3 and Column~4 then list the proper motions computed in this work. The dust masses estimated from the 1.3~mm fluxes \citep{2022A&A...663A..98T} and from those obtained from SED fitting are presented in Column~5 and Column~6. Column~7 and Column~8 give the accretion luminosity derived from our IRTF observations, or gathered from the literature. The last column remarks on our membership assessment of each $Q$ candidate. 
Four additional objects not within our $W$-band images therefore not as $Q$ candidates, but have been spectroscopically observed to look for accretion signatures are separately summarized in Table~4.

\begin{longrotatetable}
\begin{deluxetable}{ccCCCCCCCc}
\tabletypesize{\scriptsize}
\tablecaption{Youth and membership of water bearing BDs}
\tablenum{3}
\tablehead{
\colhead{No.\tablenotemark{$^a$}}  & \colhead{SpTy} & \colhead{$\mu_{\alpha}$ (CFHT-$K_S$)} & \colhead{$\mu_{\delta}$(CFHT-$K_S$)} & \colhead{$M_{Disk}$~(1.3mm)} &\colhead{$M_{Disk}$~(SED)} & \colhead{$\log$ ($ L_{\rm acc}/L_\sun$) } & \colhead{$\log$ ($ L_{\rm acc}/L_\sun$)}& \colhead{ref $L_{\rm acc} $\tablenotemark{$^\dagger$}}& \colhead{Membership}    \\
\colhead{}  & \colhead{} & \colhead{mas~yr$^{-1}$} & \colhead{mas~ yr$^{-1}$} & \colhead{$M_\earth$ } & \colhead{ $M_\earth$}   &\colhead{ (this work, IRTF)} & \colhead{(literature)} & \colhead{ }& \colhead{} }
\startdata
1 & M4 & -4.14 & -36.07 & 8.29 & 64.60 & \nodata & <-2.18 & 3 & Y \\
2 & M5 & -2.82 & -28.22 & \nodata & 1.65 & \nodata & <-3.36 & 3 & Y \\
3 & M5.5 & -11.06 & -22.84 & 0.37 & 3.16 & \nodata & <-4.4 & 4 & Y \\
4 & M5.5 & -5.33 & -28.16 & 0.2 & 0.51 & \nodata & <-4.4 & 4 & Y \\
5 & M5.75 & -19.57 & -32.15 & 2.1 & 33.96 & \nodata & <-4.02 & 3 & Y \\
6 & M6 & -14.38 & -21.56 & 0.21 & 1.03 & \nodata & \nodata & \nodata & Y \\
7 & M6 & -13.94 & -38.76 & \nodata & 0.01& \nodata & \nodata & \nodata & Y \\
8 & M6 & -19.78 & -19.46 & 12.4 & 148.52 & -1.48\pm0.13 & -1.8 & 3 & Y \\
9 & M6 & -7.41 & -38.09 & \nodata & \nodata & \nodata & \nodata & \nodata & Y \\
10 & M6.5 & -8.5 & -35.01 & 0.4 & 4.86 & \nodata & -3.65 & 3 & Y \\
11 & M7 & -11.92 & -34.03 & 0.26 & 0.05 & \nodata & <-5.82 & 3 & Y \\
12 & M8 & -12.78 & -34.8 & 1.0 & 0.01 & \nodata & <-4.1 & 4 & Y \\
13 & M8 & -4.96 & -37.59 & 0.26 & 0.16 & <-3.35 & \nodata & \nodata & Y \\
14 & M8 & -22.46 & -29.67 & 0.07 &  \nodata & \nodata & <-4.1 & 4 & Y \\
15 & M8.5 & -24.77 & -25.54 & 0.07 & \nodata & \nodata & \nodata & \nodata & Y \\
16 & M9.5 & -5.83 & -38.71 & \nodata & \nodata & \nodata & \nodata & \nodata & Y \\
17 & M8.25 & -0.54 & -25.63 & \nodata & 0.01 & <-3.60 & \nodata & \nodata & Y \\
18 & M9.75 & -11.62 & -34.33 & 0.04 & \nodata & \nodata & <-3.9 & 4 & Y \\
19 & L4 & -0.26 & -35.45 & \nodata & \nodata & \nodata & \nodata & \nodata & Y \\
20 & M8~(this work) & -10.67 & -21.21 & \nodata & 1.78 & <-2.56 & \nodata & \nodata & Y \\
21 & \nodata & -4.82 & -23.48 & \nodata & \nodata & \nodata & \nodata & \nodata & Y \\
22 & \nodata & 2.99 & -3.44 & \nodata & \nodata & \nodata & \nodata & \nodata & N \\
\enddata
\label{table:Q_param}
\end{deluxetable}

\tablenotetext{$$^a$$}{The same source ID number in Column 1 as in table~\ref{table_Qcandidates}}
\tablenotetext{$$^\dagger$$}{The reference for accretion luminosity 3: \citet{2015A&A...579A..66M}, 4: \citet{2024A&A...685A.118A} }
\end{longrotatetable}

\begin{deluxetable}{cCCcC}
\tabletypesize{\scriptsize}
\tablecaption{Additional Targets with IRTF Spectra}
\tablenum{4}
\tablehead{
\colhead{No.}  &  \colhead{$\alpha$~ (J2000)} & \colhead{$\delta$ ~ (J2000)}  &\colhead{SpT}   & \colhead{ $\log$ ($ L_{\rm acc}/L_\sun$)}  \\
\colhead{}  & \colhead{} & \colhead{ }   &\colhead{}  &\colhead{ (this work)}  }
\startdata
23 & 247.053 & -24.19322 & M6.25  & -1.83\pm 0.32 \\
24 & 247.20296 & -24.44226 & M6.25  & <-3.94 \\
25 & 246.92016 & -24.48366 & M7.5 & <-3.75 \\
26 & 246.76676 & -24.04642 & M7.25  &<-3.51  \\
\enddata
\label{table:Q_param1}
\end{deluxetable}

\section{Discussion} \label{sec:Results}

 \subsection{Masses and Ages} \label{sec:mass_and_ages}

\begin{figure}[htb!]
\includegraphics[width=\columnwidth]{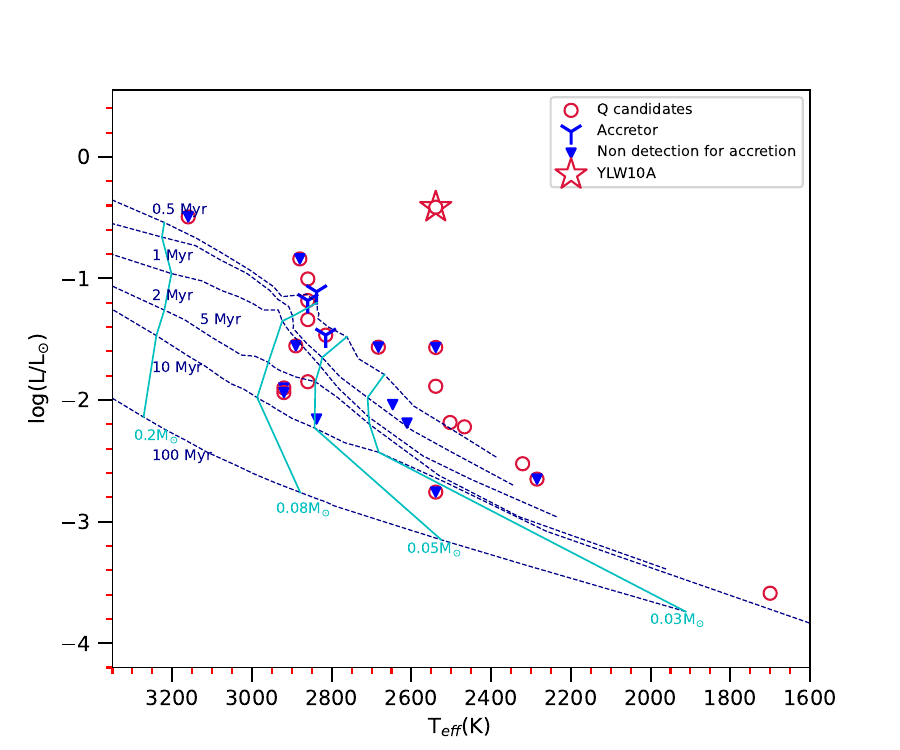} %
    \caption{Bolometric luminosity versus effective temperature.  The $Q$ candidates are symbolized in red.  Those diagnosed as accretors are marked in blue Y, and non-accretors as blue inverted triangles, as per our observations or from the literature. The mass tracks (in cyan) and isochrones (in blue) from \citet{2015A&A...577A..42B} are labelled.}
    \label{L_vs_Teff}
\end{figure}

The  mass and age are characterised by using HR diagram and isochrones. The effective temperature is converted from the spectral type adopting  the relations from \citet{2015ApJ...808...23H} for those earlier than M7, and from \citet{2015ApJ...810..158F} for later types. The bolometric luminosity is derived from $J$ mag with the bolometric correction from \citet{2015ApJ...808...23H} and \citet{2015ApJ...810..158F}.  A distance of 137.3~pc \citep{2017ApJ...834..141O} is adopted to calculate the luminosity, along with interstellar extinction derived from the SED fitting whenever available. Otherwise $A_V$ is taken from the literature. The effective temperature versus bolometric luminosity is then compared with theoretical isochrones, as depicted in Figure~\ref{L_vs_Teff}.

The majority of our sources have masses less than 0.08~$M_\sun$, and ages younger than 5~Myr; some are very low mass stars near the hydrogen burning limit, and many are BDs well within the substellar regime.  The objects with detected accretion span the mass range of 0.08~$M_\sun$ to 0.06~$M_\sun$. Most of our sources are confined near the isochrones, except YLW10A, which will be discussed later in detail. 

\subsection{Disks and Accretion} \label{sec:disk_accretion}

Disks around low mass stars and BDs provide insights into (a) planet formation (b) youth and evolutionary status (c) substellar formation pathway.

While direct observations of planet formation in BD disks are lacking, indirect evidence such as grain settling  \citep{2005Sci...310..834A} has been observed in BD disks,  indicating the potential for planet formation. With the presence of disk and estimation of disk masses, the upper limit on the mass and size of the planet to be  formed can be inferred.  In our sample, the disk dust mass obtained from the SED fitting has a median of $1.34~M_\earth$, higher than those available from continuum fluxes at 1.3~mm \citep{2019MNRAS.482..698C, 2022A&A...663A..98T},  are listed in Table~\ref{table:Q_param}). Assuming a gas to dust ratio of 100, this total disk mass corresponds to  $134~M_\earth$, or equivalently  $0.4~M_J$. It appears that  with such low disk masses, the available  material is not enough to form giant planets, even if planet formation were highly efficient. This has been also pointed out by \citet{2007MNRAS.381.1597P, 2020A&A...638A..88L} based on their simulations using core accretion models of planet formation,  and further suggesting that while the formation of Earth like planets around  BDs require disk with a few $M_J$’s, smaller Mars size planets could form in BD disks .

Our $Q$ candidates, primarily of mid-M to mid-L types, exhibit Class~II or Class~III type disks. This indicates an age less than a few million years, and evolution similar to that of low-mass stars progressing through a pre-main sequence stage. The fraction of disks in the water-bearing sample (on the basis of their SEDs) is $68\% \pm 18\%$. Previous studies have reported a similar disk frequency  of  $\sim50\%-70\%$ for young stars in Oph-core \citep{2003AJ....126.1515J, 2021ApJ...921...72M}, suggestive of common formation pathway for T-Tauri stars and BDs.  More recent studies using comparison of disk fraction of BDs and  low mass stars in other star forming regions \citet{2023JApA...44...77D, 2024MNRAS.528.5633G}  have also confirmed the BD formation as  scaled down version of low mass star formation.  The detection of accretion is limited, though not entirely absent. Out of the 17 sources observed for possible magnetospheric accretion (either with IRTF or VLT), three are found to be accretors. The remaining 14 exhibit no obvious detection, resulting in only upper limits.  The detection of accretion also depends upon factors like geometry of disk, sensitivity of our instrument for inherently weak accretors. 

In the scenario of substellar formation where a stellar embryo is ejected as a result of dynamical  and accretion is terminated, one would expect weak or  truncated disks. High fraction of disks in our water-bearing sample suggests, otherwise. Our sample albeit small indicates the presence of disks hence the notion that brown dwarf formation and evolution is a scaled-down version of star formation.

\subsection{Notes on Individual Candidates}

\begin{figure*}[htb!]
    \includegraphics[width=0.49\columnwidth]{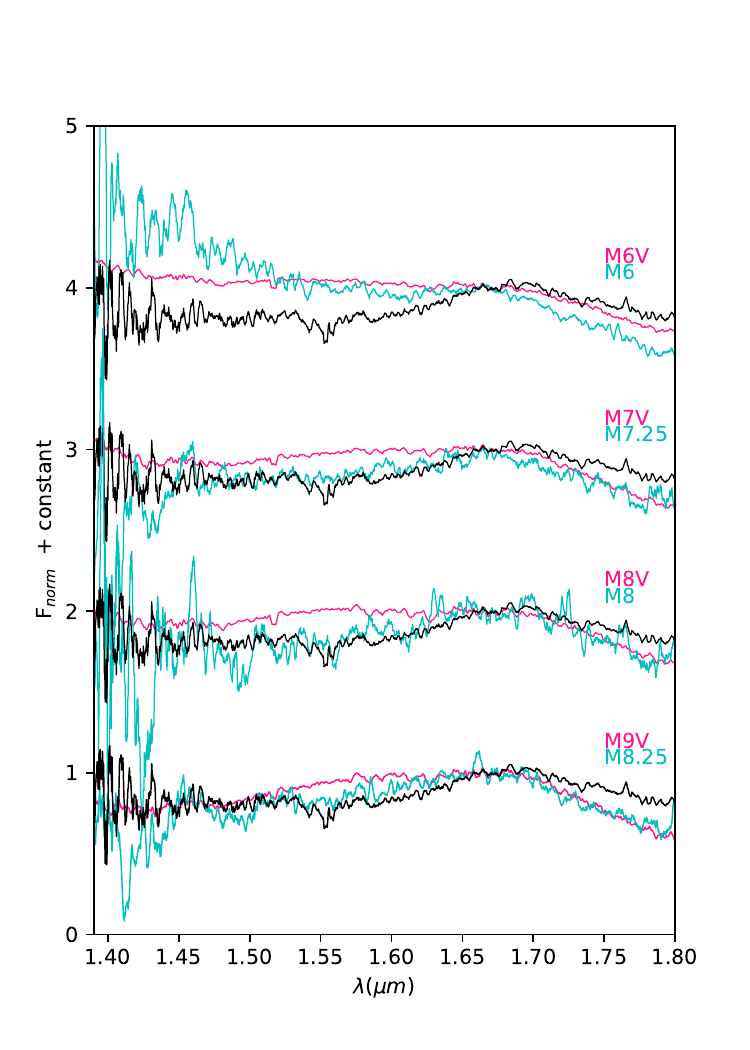} %
    \includegraphics[width=0.49\columnwidth]{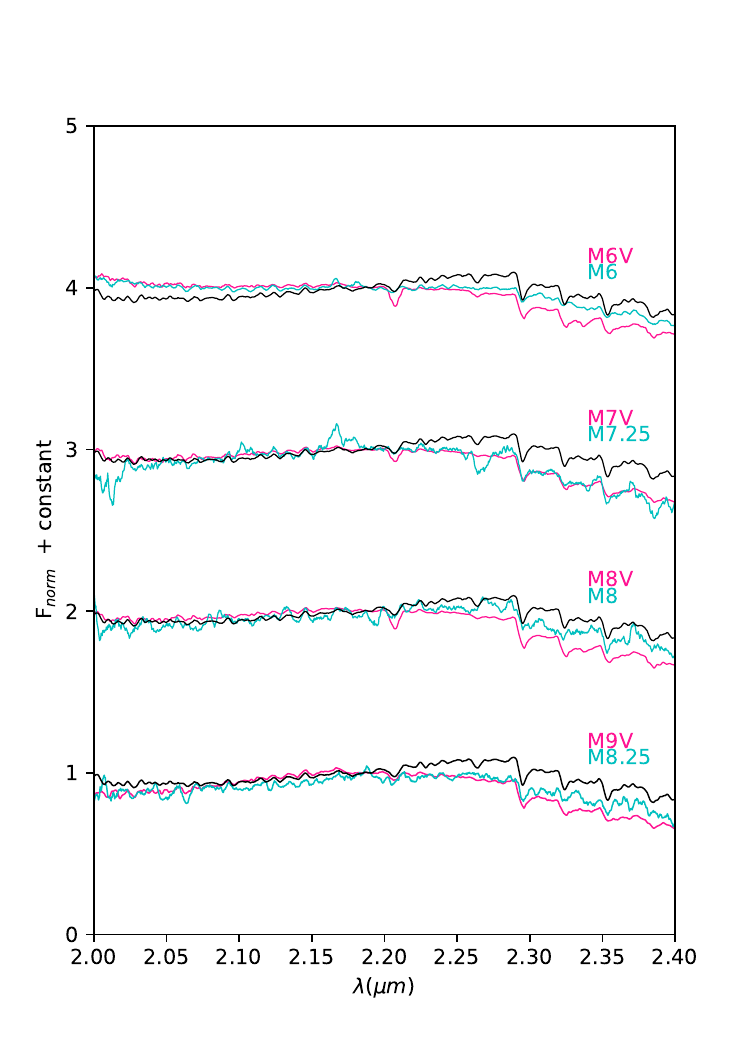}
    \caption{$H$ plus $K_s$ band extinction corrected spectrum of YLW10A, after dereddening by a visual extinction of $A_V=28.6$~mag. The spectra presented is smoothed to reduce the noise for better visual comparison. The flux is normalised at 1.67~\micron~($H$ band) and 2.19~\micron~($K_s$ band), respectively. The field dwarf sequence for mid- to late-M dwarfs is shown in magenta (from the IRTF library), the young BD sequence is shown in cyan (our observations). YLW10A appears to closely resemble the spectral type of M8 in both bands.  }
\label{Sptyping_YLW10A}
\end{figure*}

No.~20 (YLW10A) is previously classified as a young stellar object with notable infrared excess \citep{1986ApJ...304L..45Y,2015ApJS..220...11D}, as well as a Chandra X-ray source \citep{2001ApJ...557..747I}. Our W-Band data categorizes it as a $Q$ candidate, indicating water absorption at 1.45~$\mu$m. To the best of our knowledge, there is no existing spectroscopic confirmation of this source. To estimate its spectral type, we compared our IRTF spectra with the template stellar spectra of different spectral types from giants to dwarfs in the IRTF SpeX library \citep{2009ApJS..185..289R}. The templates for previously confirmed young BDs ranging from M6 to M8.5 are utilized from our observations. The spectral typing relies particularly on matching the inverted ``V'' shape to template spectra, after the best-fit extinction/reddening $Av$ is accounted. Figure~\ref{Sptyping_YLW10A} compares the target spectrum with field and young BDs for different spectral types. We deduced an M8 as the best fit, i.e., as a young BD with an extinction of $A_V = 28.6$~mag

YLW10A lies away from youngest isochrone presented in Figure~\ref{L_vs_Teff}, i.e., it is too bright as a low-mass M8 BD to be associated with this young cluster at distance of the cluster. In an alternate scenario, it could be a giant that are inherently luminous, especially if placed at a longer distance. Giants/AGBs particularly Mira variables or OH/IR stars also display a peak-like feature in the $H$~band. As AGB stars exhibit strong stellar winds, which upon cooling produce molecules and condense dust making them infrared excess sources, although the excess is typically weaker than produced by young circumstellar dust inherited from the parental molecular cloud.  Moreover, the mid-infrared light curves of six years does not indicate any significant variability for this source \citep{2021ApJ...920..132P}.   We did not find any notable Br$\gamma$ emission line in our observations that would indicate magnetospheric accretion in this source. We conclude, therefore, that based on the spectral fit, infrared excess, and X-ray detection, YLW10A is most likely to be a part of L1688A, either much younger than other stellar and substellar members or defying a proper isochrone fitting as a consequence of the copious excess emission and excessive extinction. Its position in the HR diagram and infrared excess may also result from the presence of an unresolved binary companion. Most brown dwarf binaries are quite tightly separated, and likely unresolved with seeing-limited observations. The binarity of YLW10A cannot be confirmed with existing data.

No.~8 is a $Q$-selected candidate. Our IRTF spectra show a peak in the $H$ band, confirming its nature as a young, cool, and low-gravity object. With an estimated mass of approximately 0.07~M$_\sun$, it exhibits  signs of magnetospheric accretion.  We have detected the presence of the Br$\gamma$ line, and derived an accretion luminosity of $1.48\pm0.13$ $L_\sun$. The previous estimates using Pa$\beta$ are $-1.76 L_\sun$ \citep{ 2006A&A...452..245N}, and $-1.80 L_\sun$ \citep{2015A&A...579A..66M}. While Pa$\beta$ may be optically thick and detectable due to other phenomena such as molecular outflows, the presence of Br$\gamma$, considered optically thin, provides reliable evidence of its accretion activity. 

No.~23 is an M6-type known accretor \citep{2004A&A...424..603N, 2006A&A...452..245N, 2006A&A...460..547G}. Both Pa$\beta$ and Br$\gamma$ are detected in our spectra.  We derive an accretion luminosity of $-2.68\pm1.04 L_\sun$ (using Pa$\beta$), and $-1.83\pm0.32 L_\sun$ (using Br$\gamma$). The literature values of the accretion luminosity using Pa$\beta$ are reported as $-2.87 L_\sun$ \citep{ 2006A&A...452..245N}, and $-3.08$ $L_\sun$ \citep{2015A&A...579A..66M}. Our analysis gives an estimated mass of $\sim0.08$~M$_\sun$, i.e., at the stellar and substellar mass boundary. 

\section{Conclusions}

In summary, on-off infrared imaging with a narrow-band filter centered on the 1.45~microns water absorption proves efficient in the selection of objects with cool atmospheres. Unbiased by the presence of disks, this method is particularly sensitive to selection of cool M and L dwarfs. This is in contrast to  most surveys (e.g., Spitzer, WISE etc.) targeting young populations in star-forming regions to identify disk candidates based on infrared excess.  However, W-band selection has its own limitation, as the criteria may break down in high-extinction environments. Additionally, while it is effective in identifying ultracool dwarfs, there is a degeneracy between spectral types for young BDs and field brown dwarfs. The single detection marker for water-bearing candidates, with checks on membership, leads to a reliable target list for follow-up spectroscopic confirmation. The contribution of this work follows.

\begin{enumerate} 

\item In the core of the Ophiuchus dark cloud complex, L1688\,A, the water-band imaging technique works well as 20 of the 22 candidates turn out to have spectral types ranging from M4 to L4. 

\item We conducted a membership analysis to verify association with the known young cluster by constraining their proper motions derived from deep infrared images calibrated with bright Gaia sources, and by evaluating their ages and masses.  A total of 21 out of 22 candidates were confirmed to be young members. 

\item The high fraction of disks in our sample points towards a substellar formation scenario similar to that for low-mass stars.

\item Eight relatively bright low-mass stars and brown dwarf candidates were observed spectroscopically.  In addition to affirming their cool temperatures, we confirmed accretion signatures in two very low-mass objects.

\item Spectroscopic analysis of YLW10A (No.~20) revealed an M8 spectral type. Based on the HR diagram,  YLW10A could be a potential unresolved close BD-BD binary candidate.  

\end{enumerate}

\begin{acknowledgments}
TS and WPC acknowledge the financial support for this work by the National Science and Technology Council (NSTC) grants 113-2123-M-008-004 and 113-2740-M-008-005.
\end{acknowledgments}


\end{document}